\documentclass[10pt]{IEEEtran}
\usepackage{graphicx,subfig}
\usepackage[export]{adjustbox}
\usepackage{amssymb}
\usepackage{epstopdf}
\usepackage{listings}
\usepackage{courier}
\usepackage{color}

\usepackage{parskip}
\usepackage{url}

\usepackage{enumitem}
\usepackage{datetime2}

\usepackage[utf8]{inputenc}

\title{Measurement-Based Evaluation Of Google/Apple Exposure Notification API For Proximity Detection In A Commuter Bus}
\author{Douglas J. Leith, Stephen Farrell\\
    School of Computer Science \& Statistics,\\ 
    Trinity College Dublin, Ireland\\
	15th June 2020\vspace{-5ex}
    }

\begin{document}
\maketitle

\begin{abstract}
We report on the results of a measurement study carried out on a commuter bus in Dublin, Ireland using the Google/Apple Exposure Notification (GAEN) API.  This API is likely to be widely used by Covid-19 contact tracing apps. Measurements were collected between 60 pairs of handset locations and are publicly available.  We find that the attenuation level reported by the GAEN API need not increase with distance between handsets, consistent with there being a complex radio environment inside a bus caused by the metal-rich environment.    Changing the people holding a pair of handsets, with the location of the handsets otherwise remaining unchanged, can cause variations of $\pm 10$dB  in the attenuation level reported by the GAEN API.   Applying the rule used by the Swiss Covid-19 contact tracing app to trigger an exposure notification to our bus measurements we find that no exposure notifications would have been triggered despite the fact that all pairs of handsets were within 2m of one another for at least 15 mins.  Applying an alternative threshold-based exposure notification rule can somewhat improve performance to a detection rate of 5\% when an exposure duration threshold of 15 minutes is used, increasing to 8\% when the exposure duration threshold is reduced to 10 mins.  Stratifying the data by distance between pairs of handsets indicates that there is only a weak dependence of detection rate on distance.
\end{abstract}

%%%%%%%%%%%%
\section{Introduction}
There is currently a great deal of interest in the use of mobile apps to
facilitate Covid-19 contact tracing.  This is motivated by the hope that more efficient and scalable contact
tracing might allow the lockdown measures in place in many countries to be relaxed more quickly~\cite{ferretti2020quantifying} and that these systems can help ``hedge'' against the risk of a second wave of the pandemic~\cite{secondwave}.  In early April 2020, Apple and Google formed a partnership to develop contact event detection based on Bluetooth LE~\cite{applegoogle}.  Following public launch of the Google/Apple Exposure Notification (GAEN) API on 20 May~\cite{gaenlaunch}, GAEN implementations are now installed on many people's phones and this API is likely to be widely used by national health authority contact tracing apps. 

The basic idea of a contact tracing app is that if two people carrying mobile handsets installed with the app spend significant time in close proximity to one another (e.g. spending 15 minutes within 2 metres) then the apps on their handsets will both record this contact event.  If, subsequently, one of these people is diagnosed with Covid-19 then the contact events logged on that person's handset in the recent past, e.g. over the last two weeks, are used to identify people who have been in close contact with the infected person.  These people might then be made aware of the contact and advised to self-isolate or take other appropriate precautions.   For this approach to be effective it is, of course, necessary that the app can accurately detect contact events.   

Almost all modern handsets are equipped with Bluetooth LE wireless technology and this is used by the GAEN API as the means for detecting contact events.  In general, a radio signal tends to get weaker as it gets further from the transmitter since the transmit power is spread over a greater area.   Bluetooth LE devices can be configured to transmit \emph{beacons} at regular intervals and the idea is that the signal strength with which a beacon is received provides a rough measure of the distance between transmitter and receiver.  Namely, when the received signal strength is sufficiently high then this may indicate a contact event and, conversely, when the received signal strength is sufficiently low then this may indicate that the handsets are not in close proximity.

However, the propagation of radio signals in practice is often complex, especially in indoor environments where walls, floors, ceiling, furniture etc can absorb/reflect radio waves and so change the received signal strength.   A person's body also absorbs Bluetooth LE radio signals so that the received signal strength can be substantially reduced if their body lies on the path between the transmitter and receiver.

A key difficulty in evaluating proximity detection accuracy in real-world settings is establishing ground truth i.e. recording when contact events actually happened.  This ground truth is needed so that the contact events flagged by a contact tracing app can be compared against the actual contact events and so allow the accuracy of the app at detecting contact events to be assessed.  Following \cite{techreport2020}, to address this we adopt a scenario-based approach in which the ground truth is clear (to within experimental error).   The disadvantage is that this limits study to fairly simple, well structured scenarios.  However, by selecting scenarios that aim to capture some of the key elements in common activities we can still gain useful insight into the real-world performance of Bluetooth LE received signal strength for proximity detection.

In this paper we report on the results of a measurement study carried out on a commuter bus in Dublin, Ireland.  The bus is of a standard double-decker design widely used in Ireland and the UK. Measurements were collected between 60 pairs of handset locations and we have made those publicly available~\cite{bus_dataset}.  Contact tracing apps will likely be used as an adjunct to existing manual contact tracing and test systems.  These manual systems can usually readily identify the people with whom an infected person share accommodation and with work colleagues with whom the infected person is in regular contact.  More difficult is to identify people travelling on public transport with whom an infected person has been in contact, since the identities of these people are usually not known to the infected person and are generally not otherwise recorded.   Public transport is therefore potentially an important use case where effective contact tracing apps may be of significant assistance in infection control.

In summary, our measurements indicate that radio signal propagation is highly complex within the bus used, and in particular the attenuation levels reported by the GAEN API need not increase with distance.   This is likely due to reflections from the metal walls, floor and ceiling within the bus, metal being known to be a strong reflector of radio signals.   We observe that changing the people holding a pair of handsets, with the location of the handsets otherwise remaining unchanged, can cause variations of $\pm 10$dB  in the attenuation level reported by the GAEN API.    

The GAEN API is intended for use by health authority Covid-19 contact tracing apps.  Switzerland, for example, deployed a Covid-19 contact tracing app based on the GAEN API on 26 May 2020~\cite{swiss_bbc}.   Applying the rule~\cite{swiss} which that app uses to trigger an exposure notification to our bus measurements we find that no exposure notifications would have been triggered.  This is despite the fact that in our measurements all pairs of handsets were within 2m of one another for at least 15 minutes (the case requirement of the Swiss app, and others).   We also applied an alternative threshold rule for triggering exposure notifications to our dataset, similar to current GAEN guidelines.    We find that attenuation level thresholds of up to 70dB (a high level, that previous measurements indicate would be likely to trigger false alarms in outside environments~\cite{techreport2020}) the detection rate is at most 5\% when an exposure duration threshold of 15 minutes is used, increasing to 8\% when the exposure duration threshold is reduced to 10 mins.  Stratifying the data by distance between pairs of handsets indicates that there is only a weak dependence of detection rate on distance, consistent with the complex nature of the radio environment already noted.

%%%%%%%%%%%%
\section{Preliminaries}
%%%%%%%%%%%%%%%
\subsection{Brief Overview of Bluetooth LE}
Bluetooth Low Energy (LE) was standardised in 2010.  The low energy moniker refers to the reduced drain on the device battery compared to the older Bluetooth Classic technology. The first mobile handsets using Bluetooth LE appeared in 2011-12 (e.g. the iPhone 4S) and today almost all modern handsets come equipped with it.  

Bluetooth LE operates in the same 2.4GHz unlicensed radio band as WiFi and other devices (including microwave ovens).   Bluetooth LE devices advertise their presence by periodically (typically once per second) broadcasting short beacon messages.  To mitigate the effects of interference from other users of the 2.4GHz band beacons are broadcast on three widely spaced radio channels.   

%https://www.ti.com/lit/an/swra475a/swra475a.pdf
Each beacon essentially consists of a short fixed \emph{preamble}, followed by a small beacon \emph{payload}.  The payload contains an identifier of the device making the broadcast (in modern devices this identifier is usually randomised and changes frequently to improve privacy) plus a short message (generally up to 31 bytes long).  This message is typically used to indicate that the beacon is associated with a particular app or service, e.g. to associate it with a contact tracing app.

A device equipped with a Bluetooth LE receiver scans the three beacon radio channels listening for beacon transmissions.   When the start of a transmission is detected the receiver uses the fact that the beacon preamble is fixed and known to fine tune the radio receiver to the incoming signal.  As part of this fine tuning process a received signal strength indicator (RSSI) is output, which is an estimate of the radio power in the received signal.    If the received signal strength is too weak either the transmission is simply not noticed or this fine-tuning process fails.  Typically this occurs when the received signal strength is below around -90dB to -100dB (the noise floor of the receiver).   Upon successful fine-tuning of the receiver the payload of the beacon is decoded and passed up to the operating system and then on to relevant apps.

The received signal strength is affected by the transmit power used by the device broadcasting the beacon.   Bluetooth LE devices generally use a relatively low transmit power (to save on battery drain) and a rough guideline is that beacons cannot be decoded at distances beyond about 5-10 metres from the transmitter.  In practice the received signal strength is, however, also greatly affected by the way in which the radio signal propagates from transmitter to receiver.  In general the radio signal gets weaker as it travels further since the transmit power is spread over a greater area.  However, many complex effects can be superimposed upon this basic behaviour.  In particular, obstacles lying on the path between the transmitter and receiver (furniture, walls etc) can absorb and/or reflect the radio signal and cause it to be received with higher or lower signal strength.  A person's body also absorbs radio signals in the 2.4 GHz band and so the received signal strength can be substantially reduced if their body lies on the path between the transmitter and receiver.   In indoor environments walls, floors and ceilings can reflect radio signals even when they are not on the direct path between transmitter and receiver, and so increase or decrease the received signal strength.    See, for example, ~\cite{techreport2020} for measurements illustrating such effects in real environments.

Metal, in particular, strongly reflects radio waves and this can be an important factor in radio propagation in environments with a lot of metal.  In buses the walls, floor and ceiling are mainly metal and the seats often contain metal parts.  We can therefore expect that radio propagation in these environments will be complex, and in particular due to reflections the signal strength may not decrease as quickly with distance as in other environments e.g. see~\cite{trains2009,trains2019}.

%%%%%%%%%%%%
\subsection{GAEN API}
Use of the GAEN API is limited by Google to health authority apps or to handsets registered with a limited set of gmail accounts included on a whitelist maintained by Google.   The GAEN system is closed-source, and the available documentation provides few details as to its internal operation.  The main focus of the GAEN system documentation is instead on the specification two interfaces, which we summarise below.  See the GAEN documentation~\cite{gaen_docs} and the recent independent analysis in~\cite{duediligencereport2020} for further details.

\subsubsection{Bluetooth Beacon Format}
The first interface specified is the format to be used for Bluetooth LE beacons to ensure interoperability between handsets, in particular between handsets running Apple's iOS operating system and handsets running Google's Android operating system.    In summary, each handset generates a random \emph{Temporary Exposure Key} (TEK) once a day.   This TEK is then used to generate a sequence of \emph{Rolling Proximity Identifiers} (RPIs), approximately one for each 10 minute interval during the day (so around 144 RPIs are generated).   The GAEN system running on a handset transmits beacons roughly every 250ms.   Each beacon contains the current RPI value.  Approximately every 10 minutes the beacons are updated to transmit the next RPI value.   By constantly changing the content of beacons in this way the privacy of the system is improved.   In addition to the RPI each beacon also carries encrypted \emph{metadata} containing the wireless transmit power level used. Although beacons are emitted roughly every 250ms, on the receiving side, devices only scan for beacons roughly every 4 minutes.

\subsubsection{Query Interface}
The second interface is between the GAEN system running on a handset and apps running on the same handset.   This interface allows apps to submit a request that includes an Exposure Configuration data structure to the GAEN system~\cite{gaen_docs}.   The Exposure Configuration data structure allows specification of the TEK to be queried, the start time and duration of the interval of interest (specified in 10 minute intervals since 1st Jan 1970) and a low and high attenuation threshold (specified in dB).  The GAEN system responds with one or more Exposure Information data structures that report an exposure duration (field durationMinutes) and an array with three \emph{atttenuation duration} values, giving the duration (in minutes) that the attenuation level is below the low threshold, the duration the attenuation level is between the low and high thresholds and the duration above the high threshold.   It is also possible to query for an Exposure Summary response, but we did not make use of this since the relevant information that this contains can be derived from the Exposure Information reports.   

%% https://developer.apple.com/documentation/exposurenotification/enexposureconfiguration does say that
%% the attenuation is tx-rssi though.
The GAEN documentation does not precisely state how the attenuation level is calculated, nor does it give details as to how the attenuation duration is calculated.  The analysis in~\cite{duediligencereport2020} deduces that the attenuation level is  calculated as $P_{TX}-P_{RX}$, where $P_{TX}$ is the transmit power level sent in the beacon metadata and $P_{RX}$ is given by the filtered RSSI plus a calibration value.   

We also note the same analysis indicates that the GAEN API uses a filtered RSSI value when calculating attenuation levels and durations~\cite{duediligencereport2020}.  Namely, for Google Pixel 2 handsets (and others) the RSSI is recorded only from beacons transmitted on one of the three radio channels used by Bluetooth LE for transmitting beacons.   

%%%%%%
\subsection{Android Bluetooth LE Scanner API }
The Android operating system includes a standard Bluetooth LE Scanner API.  Any app with the appropriate permissions can access this API, unlike the GAEN API.   The scanner API can be configured to report an RSSI value for all beacons received by a handset.

%\subsubsection{Mapping From RSSI To Attenuation}
%Note that while the GAEN system measures the RSSI of beacons received the query interface reports {attenuation} durations below/within/above the low and high attenuation levels specified in the Exposure Configuration data structure.   The mapping from RSSI to attenuation level is not specified in the GAEN API documentation but analysis indicates that attenuation is calculated as $P_{TX}-P_{RX}$ where $P_{TX}$ is the transmit power level sent in the beacon metadata (-17dB for Google Pixel 2 handsets) and $P_{RX}$ is given by the filtered RSSI plus a calibration value which is -4dB for Google Pixel 2 handsets, see~\cite{duediligencereport2020}.

%%%%%%%%%%%%
\section{Methodology}

%%%%%%%%%%%%
\subsection{Experimental Protocol}
\begin{figure}
\centering
%\subfloat[Upstairs.  Person 6 (not shown) is seated in front seat, 3.5 from person 1.]{
%\includegraphics[width=0.45\columnwidth]{bus_layout_upstairs}
%}
\subfloat[]{
\includegraphics[width=0.39\columnwidth,valign=t]{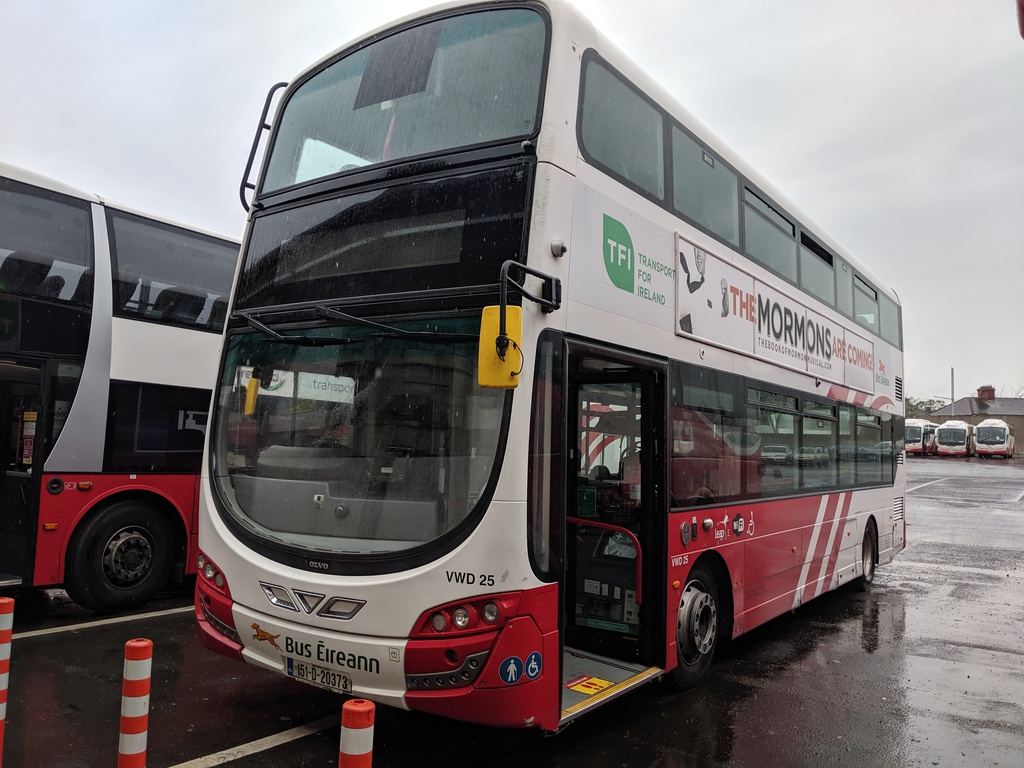}
}
\qquad
\subfloat[]{
\includegraphics[width=0.31\columnwidth,valign=t]{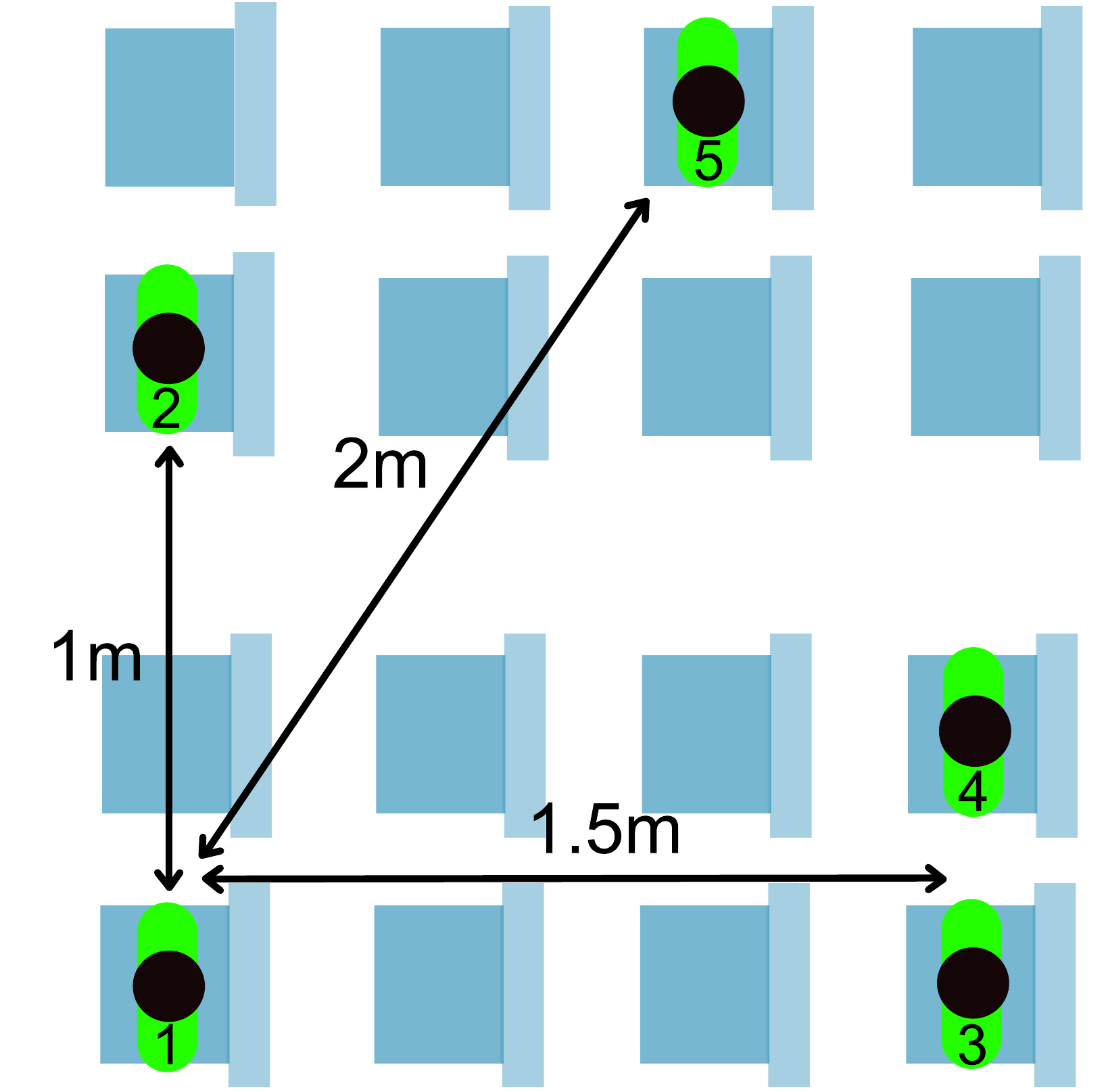}
}
\caption{(a) Bus on which measurements were collected. (b) Relative positions of participants during tests.}\label{fig:one}
\end{figure}

Our experimental measurements were collected on a standard double-decker bus used to carry commuters in Dublin, Ireland, see Figure \ref{fig:one}(a).  We recruited five participants and gave each of them Google Pixel 2 handsets.  We asked them to sit in the relative positions shown in Figure \ref{fig:one}(b).  This positioning aims to mimic passengers respecting the relaxed social distancing rules likely during easing of lockdown (where a minimum of 2m distancing is  mandated).   % It includes two people sitting together and two people sitting 2m apart to provide baseline references.
Each experiment is 15 minutes duration giving around 3 scans by the GAEN API when scans are made every 4 mins.  A Wifi hotspot was set up on the bus and the participants were asked to hold the handset in their hand and use it for normal commuter activities such as browsing the internet.     

The first experiment was carried out on the lower deck of the bus, participants were then asked to switch seats (they chose seats themselves) and a second 15 minute experiment run.    With a mix of three participants from the first two experiments and two new participants these experiments were then repeated on the upper deck of the bus.   

Each handset had the GAEN API and a modified version of the Google exemplar Exposure Notification app~\cite{modifiedENapp} installed, and was registered to a gmail user included on the Google GAEN whitelist so as to allow use of the GAEN API by the Exposure Notification app.  Each handset also had a GAENAdvertiser app developed by the authors installed.   This app implements the transmitter side of the GAEN API and allowed us to control the TEK used and also to start/stop the broadcasting of Bluetooth LE beacons.

At the start of each 15 minute experiment participants were asked to configure the GAENAdvertiser app  with a new TEK and then to instruct the app to start broadcasting GAEN beacons.  At the end of the experiment participants instructed the GAENAdvertiser to stop broadcasting beacons.   In this way a unique TEK is associated with each handset in each experiment, and these can be used to query GAEN API to obtain separate exposure information reports for each handset in each experiment.  

Following all four experiments the handsets were collected, the TEKs used by each handset extracted and the GAEN API on each then queried for exposure information relating to the TEKs of the other handsets.   At the start of the fourth experiment one participant exited the test.   In total, therefore, from these experiments we collected GAEN API reports on Bluetooth LE beacon transmissions between 120 pairs of handset locations.  Since the radio transmission path between two handsets symmetric, this yields 60 unique pairs of handset locations.   This measurement data is publicly available~\cite{bus_dataset}.

To provide baseline data on the radio propagation environment we also used the standard Android Bluetooth LE scanner API to collect measurements of RSSI as the distance was varied between two Google Pixel 2 handsets placed at a height of approximately 0.5m (about the same height as the bus seating) in the centre aisle of the upper deck of the bus.  

%%%%%%%%%%%%
\subsection{Ethical Approval}
The experimental protocol was reviewed and approved by the Ethics Committee of the School of Computer Science and Statistics, Trinity College Dublin.   The ethics application reference number is 20200503.

%%%%%%%%%%%%
\subsection{Hardware \& Software Used}
We used five Google Pixel 2 handsets running GAEN API version 202490002\footnote{As reported in the \emph{Settings-COVID 19 Notifications} handset display.}.    In a small number of measurements we also used a Samsung Galaxy A10, and we indicate when this is the case.

We used a version of the Google exemplar Exposure Notification app modified to allow us to query the GAEN API over USB using a python script (the source code for the modified app is available on github \cite{modifiedENapp}). 

In addition we also wrote our own GAENAdvertiser app that implements the Bluetooth LE transmitter side of the GAEN API.   GAENAdvertiser allows us to control the TEK, and in particular reset it to a new value at the start of each experiment.  In effect, resetting the TEK makes the handset appear as a new device from the point of view of the GAEN API, and so this allows us to easily collect clean data (the GAEN API otherwise only resets the TEK on a handset once per day).   We carried out extensive tests running GAENAdvertiser and the GAEN API on the same device to confirm that under a wide range of conditions the responses of the GAEN API on a second receiver handset were the same for beacons from GAENAdvertiser and the GAEN API, see~\cite{duediligencereport2020} for further details.   

GAENAdvertiser is open source and can be obtained by contacting the authors (we have not made it publicly available, however, since it can be used to facilitate a known replay attack against the GAEN API~\cite{replay}).

%%%%%%%%%%%%
\subsection{Querying the GAEN API}

%Querying the GAEN API on a handset involves submitting an Exposure Configuration data structure to the API.  The response is one or more Exposure Information data structures~\cite{gaen_docs}.   Note that there it is also possible to query for an Exposure Summary response, but we did not make use of this since the relevant information that this contains can be derived from the Exposure Information reports.   
%
%The Exposure Configuration data structure allows specification of the TEK to be queried, the start time and duration of the interval of interest (specified in 10 minutes intervals since 1st Jan 1970) and a low and high attenuation threshold (specified in dB).  An Exposure Information data structure reports an exposure duration (field durationMinutes) and an array with three atttenuation duration values, giving the attenuation duration (in minutes) below the low threshold, the  attenuation duration between the low and high thresholds and the attenuation duration above the high threshold.

Repeated queries were made to the GAEN API holding the low threshold constant at 48dB (which is lower than any attenuation value seen in our experiments), and varying the high threshold from 49dB to 100dB (in 1dB steps up to 80dB, then in 5dB steps since noise tends to be higher at higher attenuation levels).   By differencing this sequence of reports we infer the attenuation duration at each individual attenuation level from 48dB through to 100dB.   

%%%%%%%%%%%%
%\subsection{Visualising GAEN API Reports}

We present the attenuation duration data obtained in this way using a coloured heatmap.  We split the range of attenuation values shown on the y-axis into 2dB bins, i.e. 70-72dB, 72-74dB and so on, up to 80dB when 5dB bins are thereafter used since the data is noisier at these low signal levels.  Within each bin the colour indicates the percentage of the total duration reported by the GAEN API that was spent in that bin, e.g bright green indicates that more than 90\% of the time was spent in that bin.  The mapping from colours to percentages is shown on the righthand side of the plot.  Bins with no entries (i.e. with duration zero) are left blank.  Where appropriate we also include a solid line in plots that indicates the average attenuation level at each transmit power level (the average is calculated by weighting each attenuation level by the duration at that level and then summing over all attenuation levels).

\begin{figure}
\centering
\subfloat[]{
\includegraphics[width=0.45\columnwidth,valign=t]{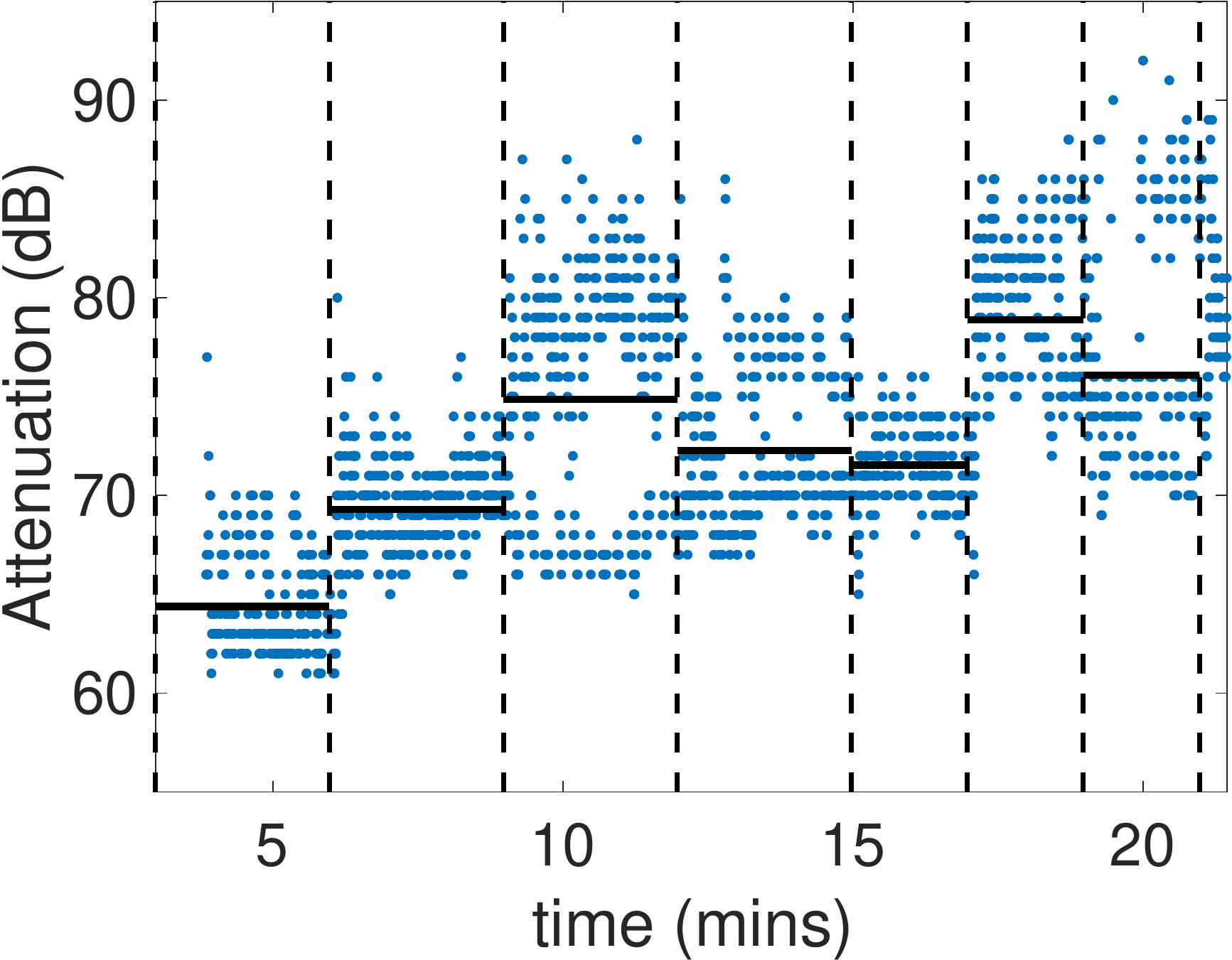}
}
\subfloat[]{
\includegraphics[width=0.39\columnwidth,valign=t]{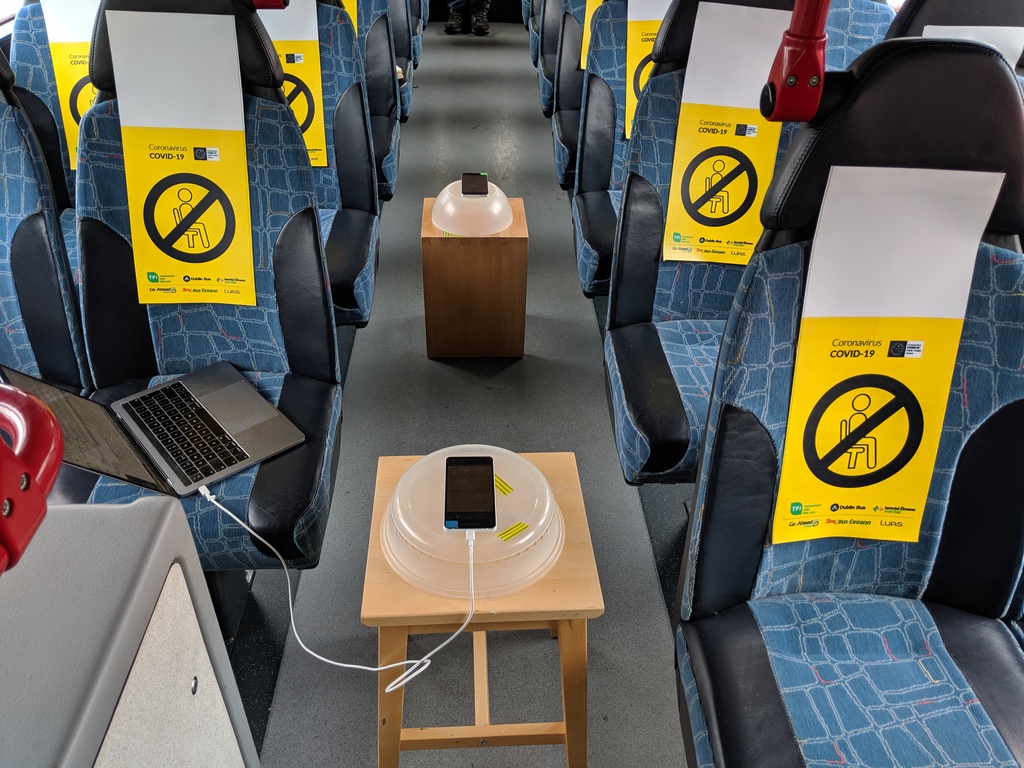}
}
\caption{(a) Measurements of attenuation between two handsets as the distance between them is varied along the centre aisle in the upper deck of the bus, (b) shows the setup used.  The vertical dashed lines indicate when the distance between the handsets was changed, starting at 0.5m and then increasing by 0.5m at each step.  The solid horizontal lines indicate the mean attenuation level at each distance.  Measurements taken using the standard Android Bluetooth LE scanner API.}\label{fig:rssi}
\end{figure}

%%%%%%%%%%%%%%
\section{Results}

\begin{figure}[t]
\centering
\subfloat[]{
\includegraphics[width=0.45\columnwidth]{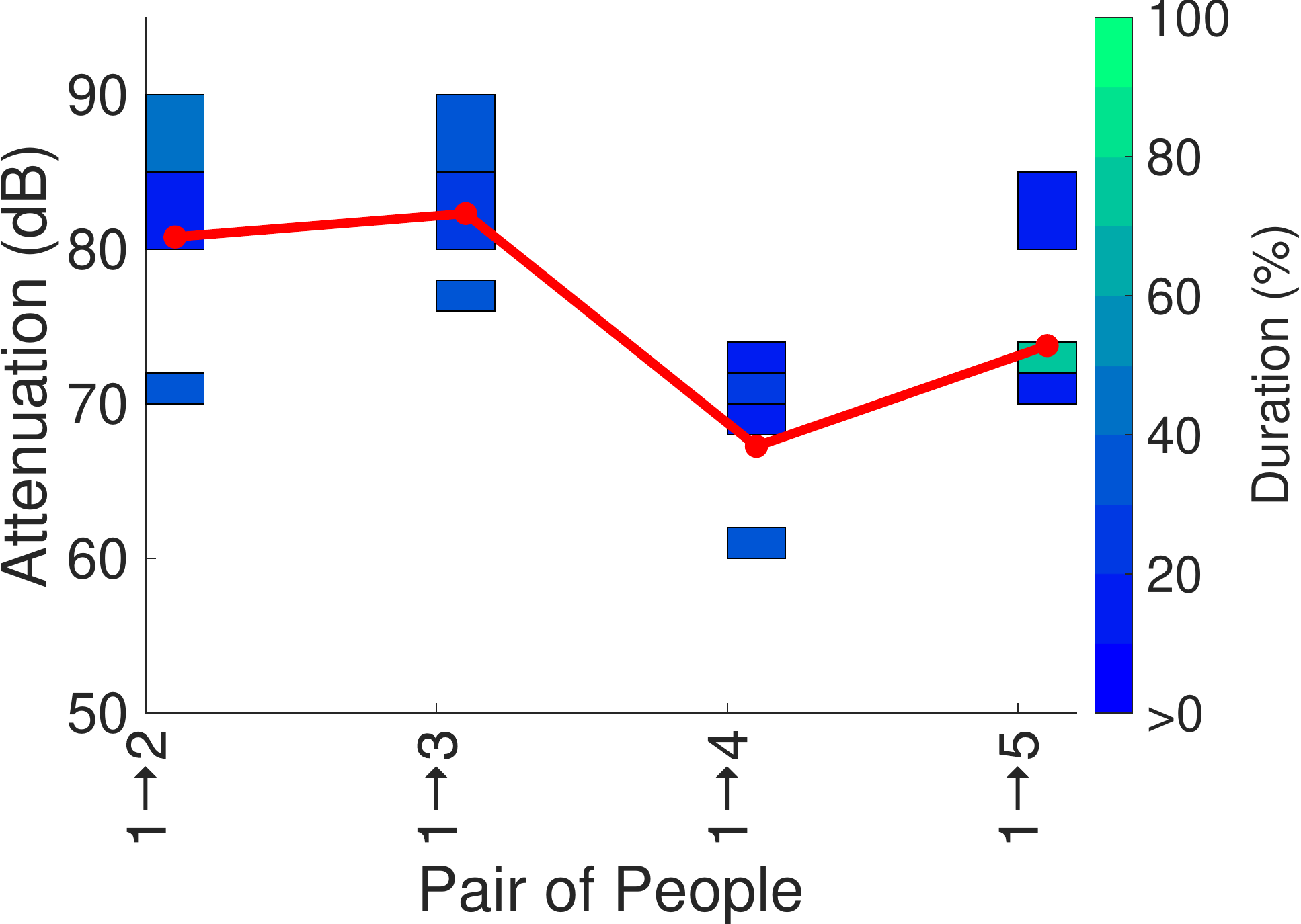}
}
\subfloat[]{
\includegraphics[width=0.45\columnwidth]{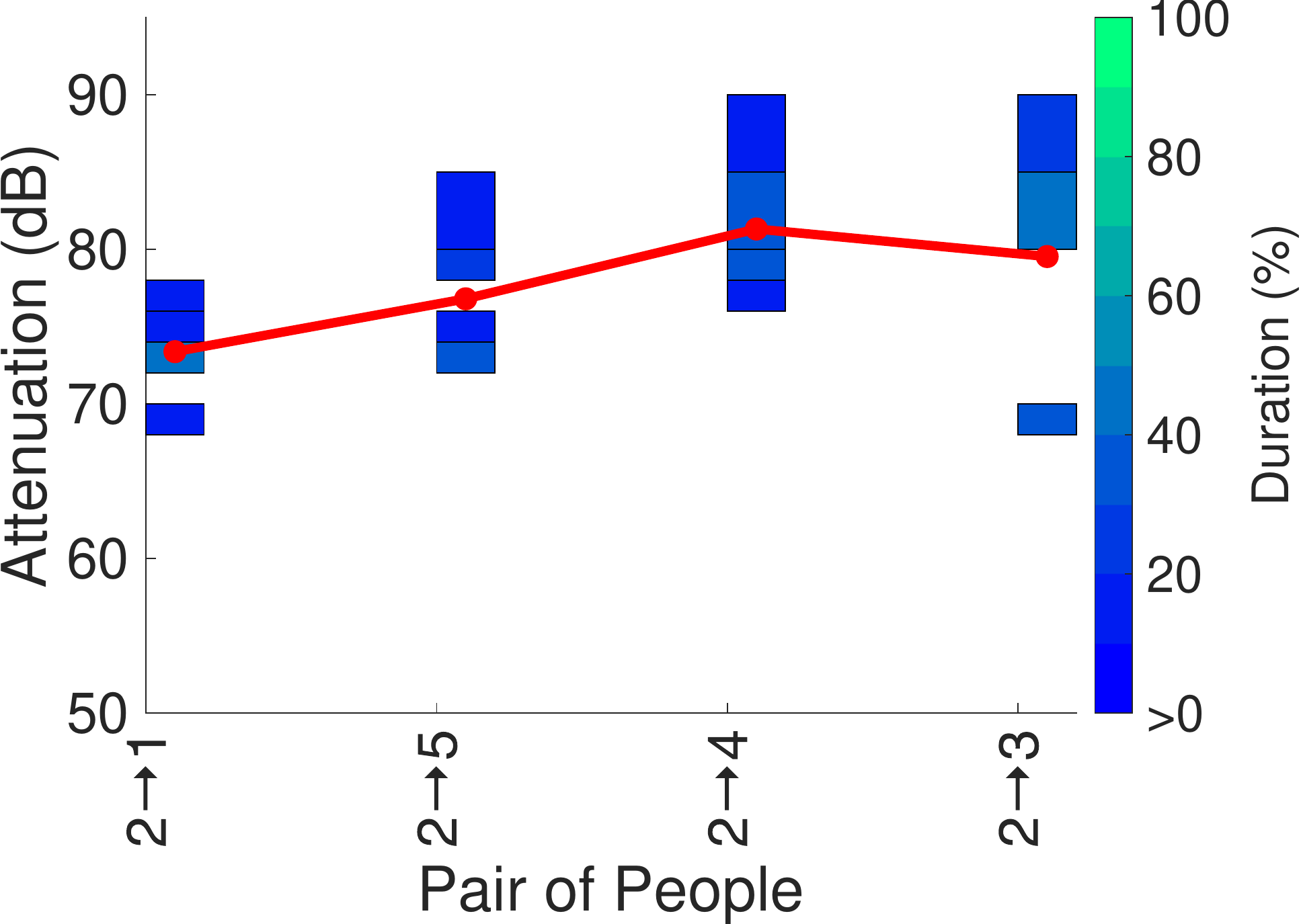}
}\\
\subfloat[]{
\includegraphics[width=0.45\columnwidth]{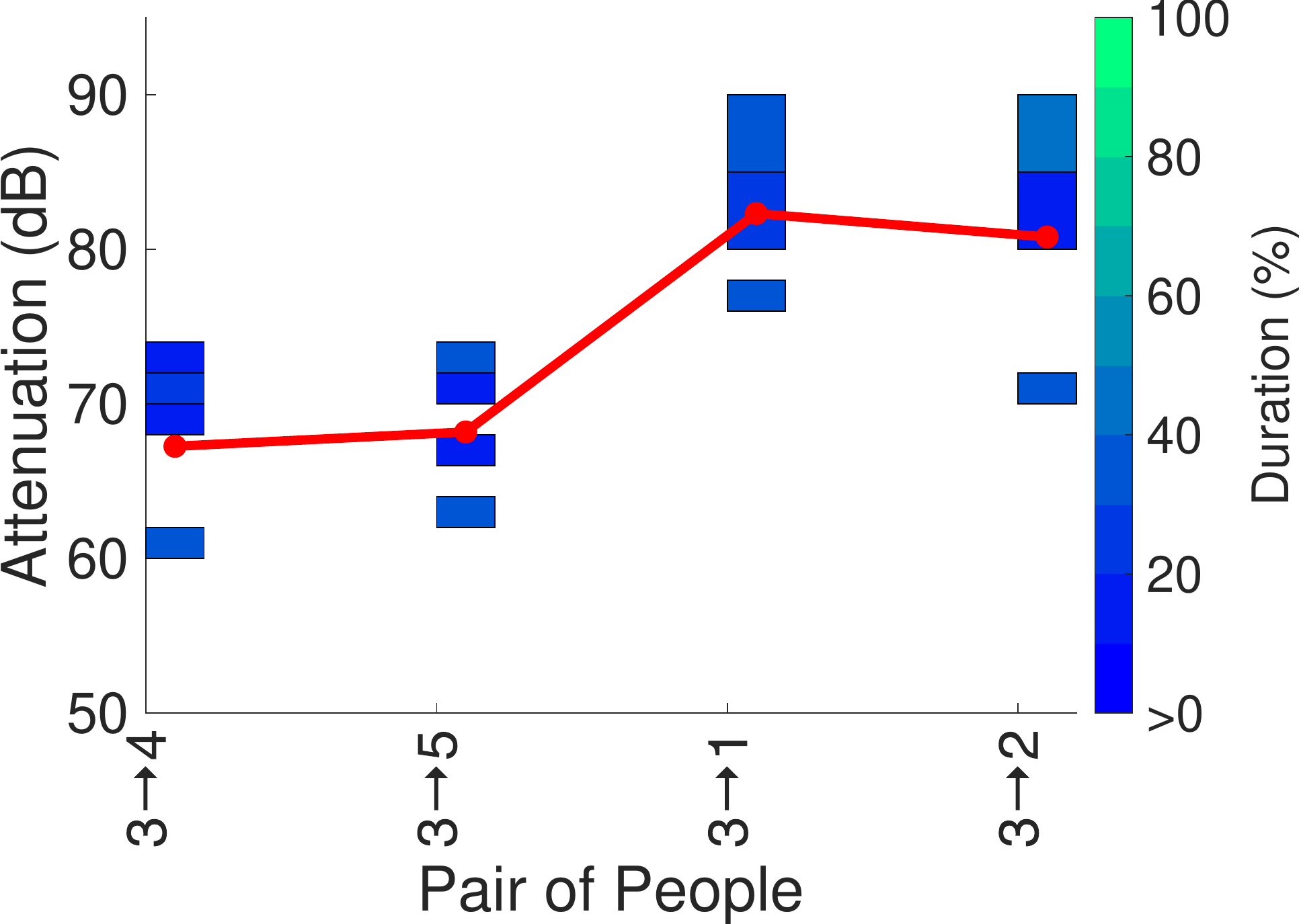}
}
\subfloat[]{
\includegraphics[width=0.45\columnwidth]{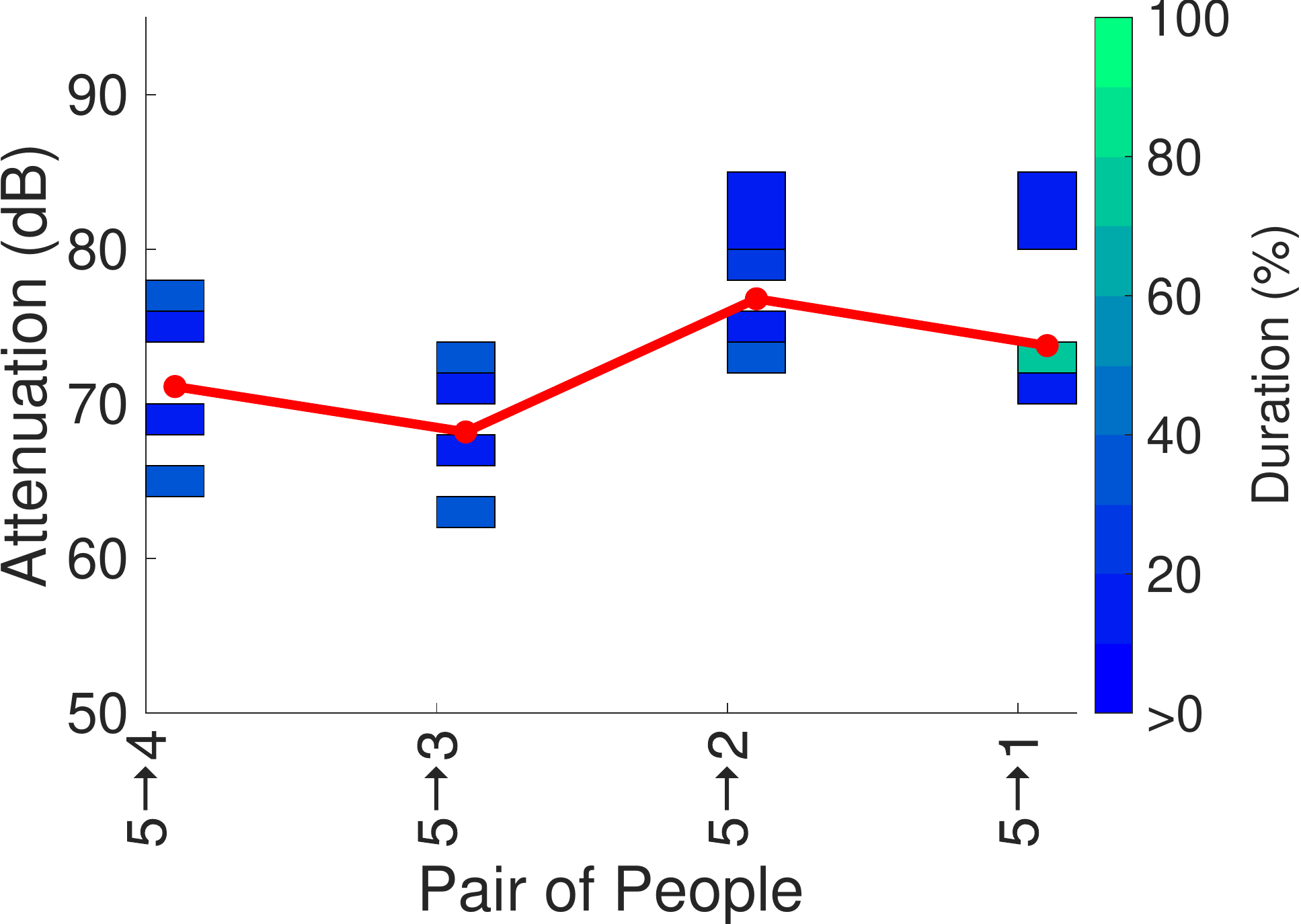}
}
\caption{Attenuation durations reported by GAEN API on completion of the first test on the lower deck of the bus.}\label{fig:test1}
\end{figure}

\begin{figure}
\centering
\subfloat[]{
\includegraphics[width=0.45\columnwidth]{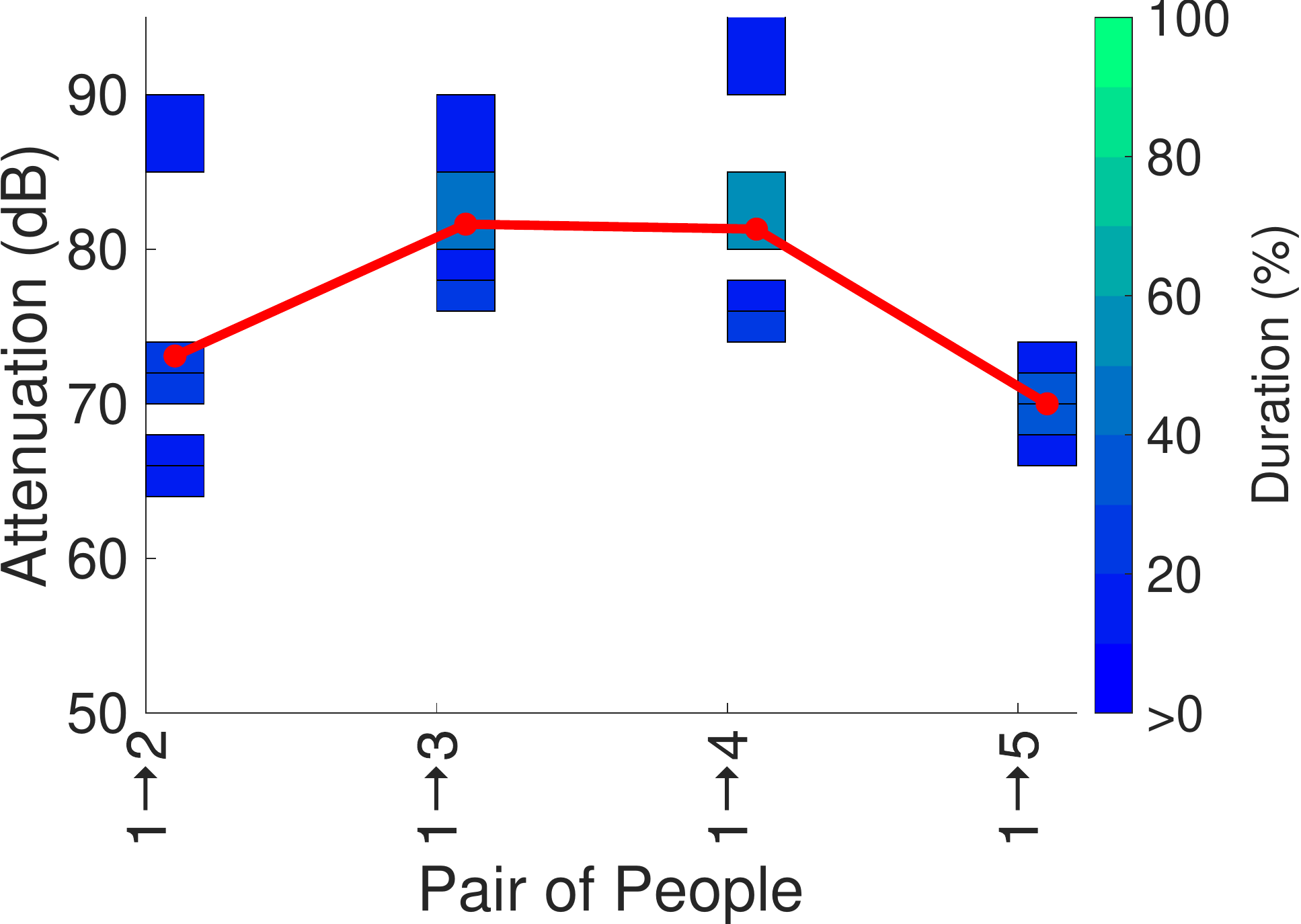}
}
\subfloat[]{
\includegraphics[width=0.45\columnwidth]{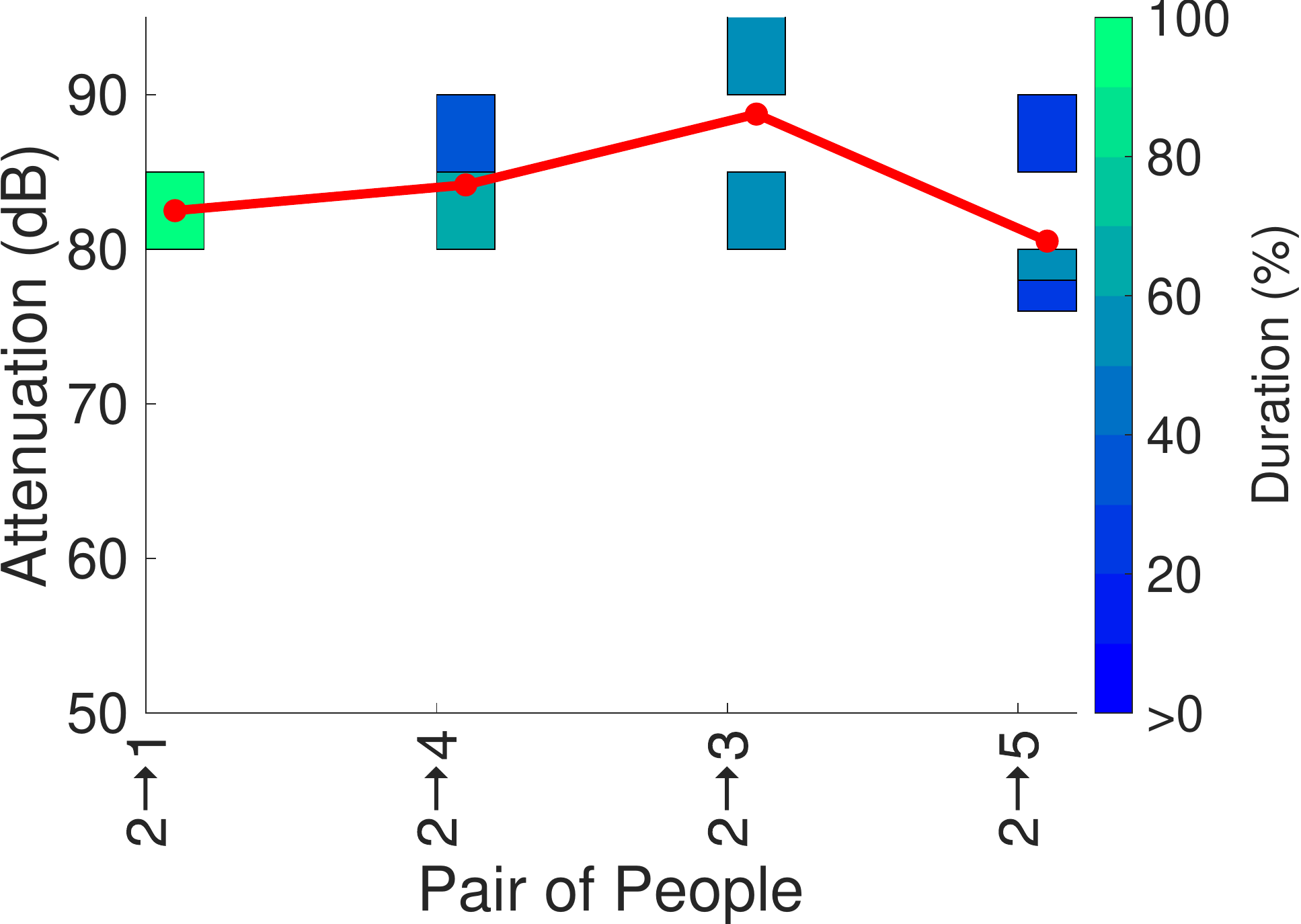}
}\\
\subfloat[]{
\includegraphics[width=0.45\columnwidth]{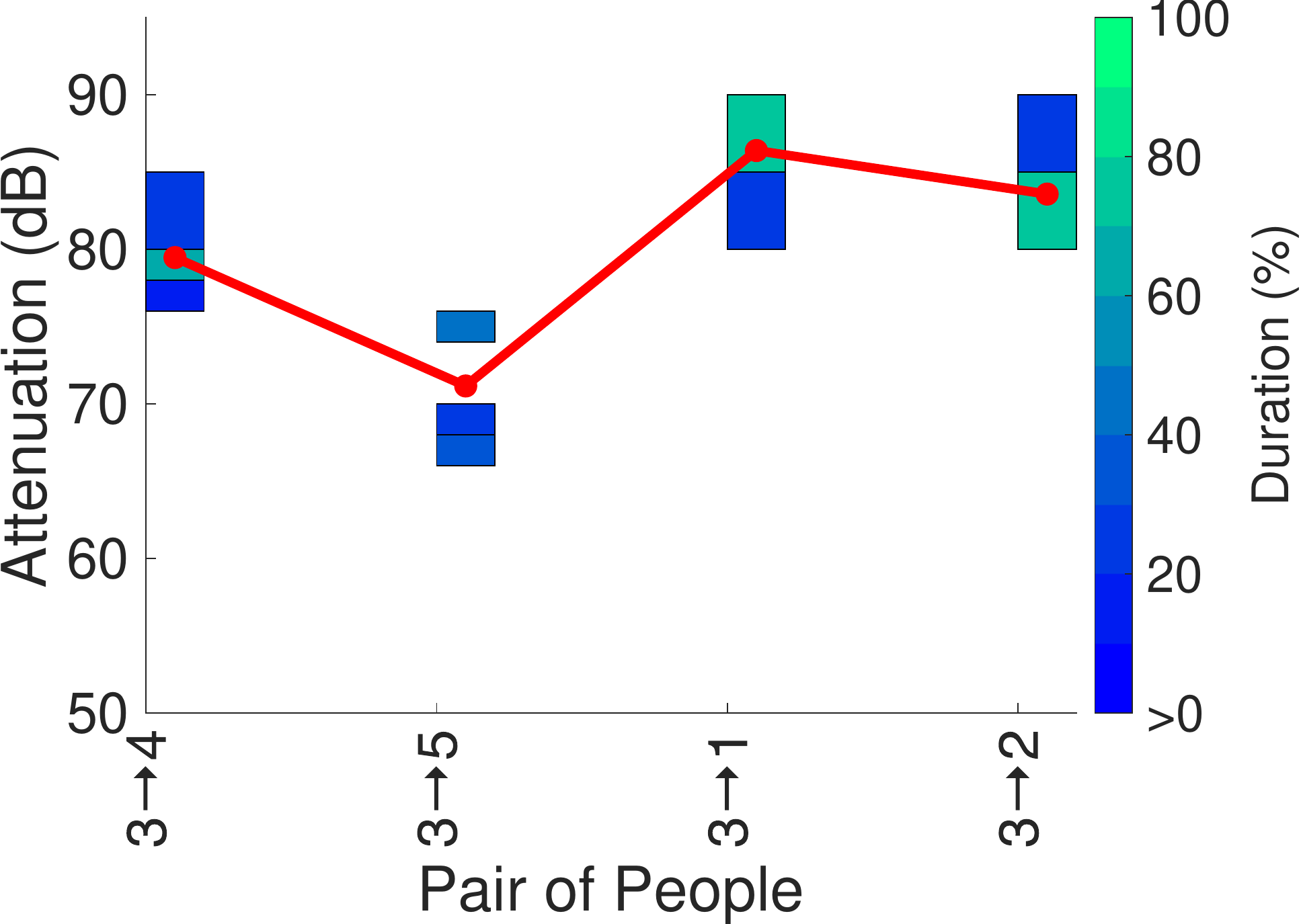}
}
\subfloat[]{
\includegraphics[width=0.45\columnwidth]{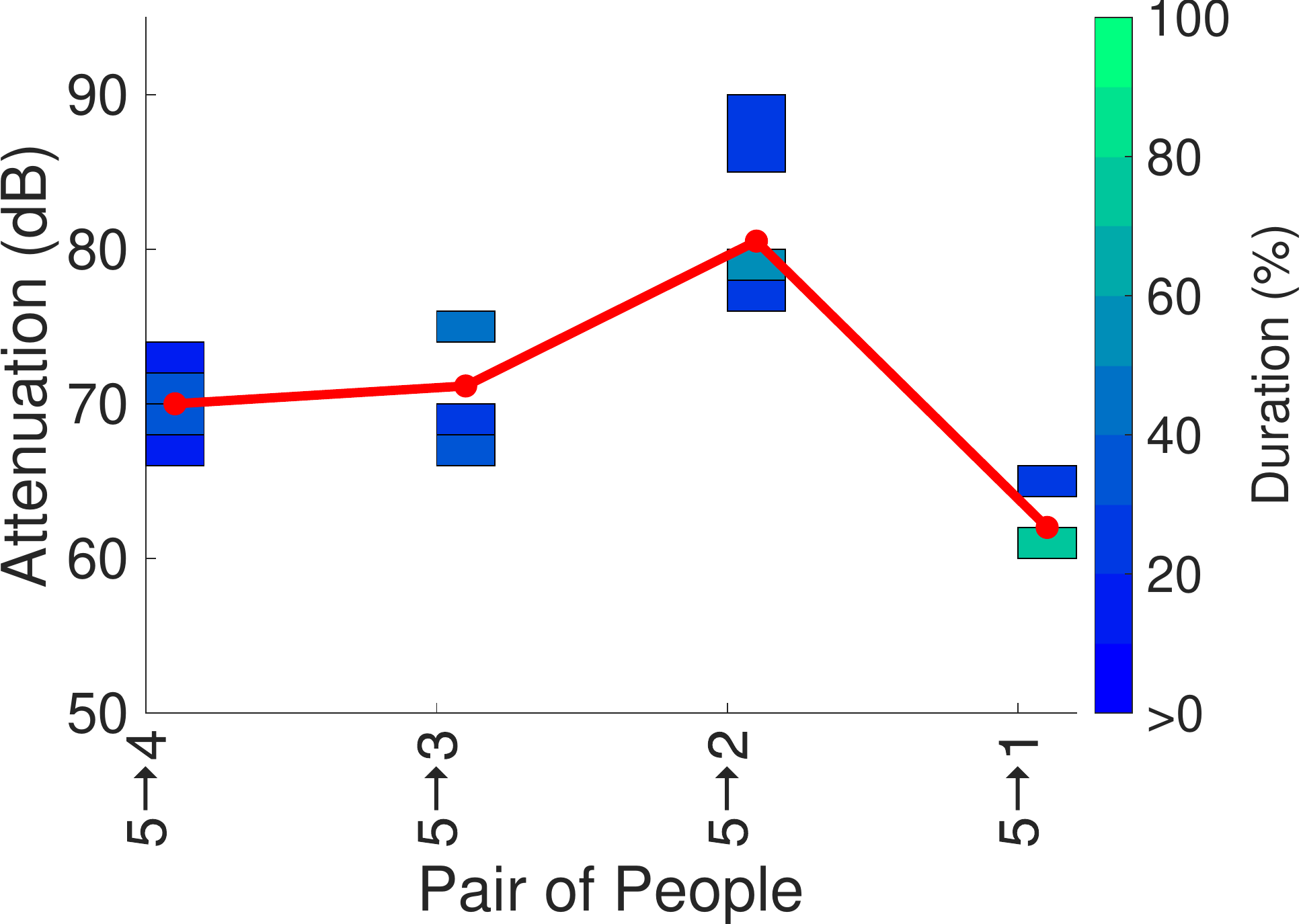}
}
\caption{Attenuation durations reported by GAEN API on completion of the second lower deck test (with the same participants as in the first test, but with their seating positions swapped about).  In (d) person 5 is using a Samsung Galaxy A10 rather than a Google Pixel 2.}\label{fig:test2}
\end{figure}

\begin{figure}[t]
\centering
\subfloat[]{
\includegraphics[width=0.45\columnwidth]{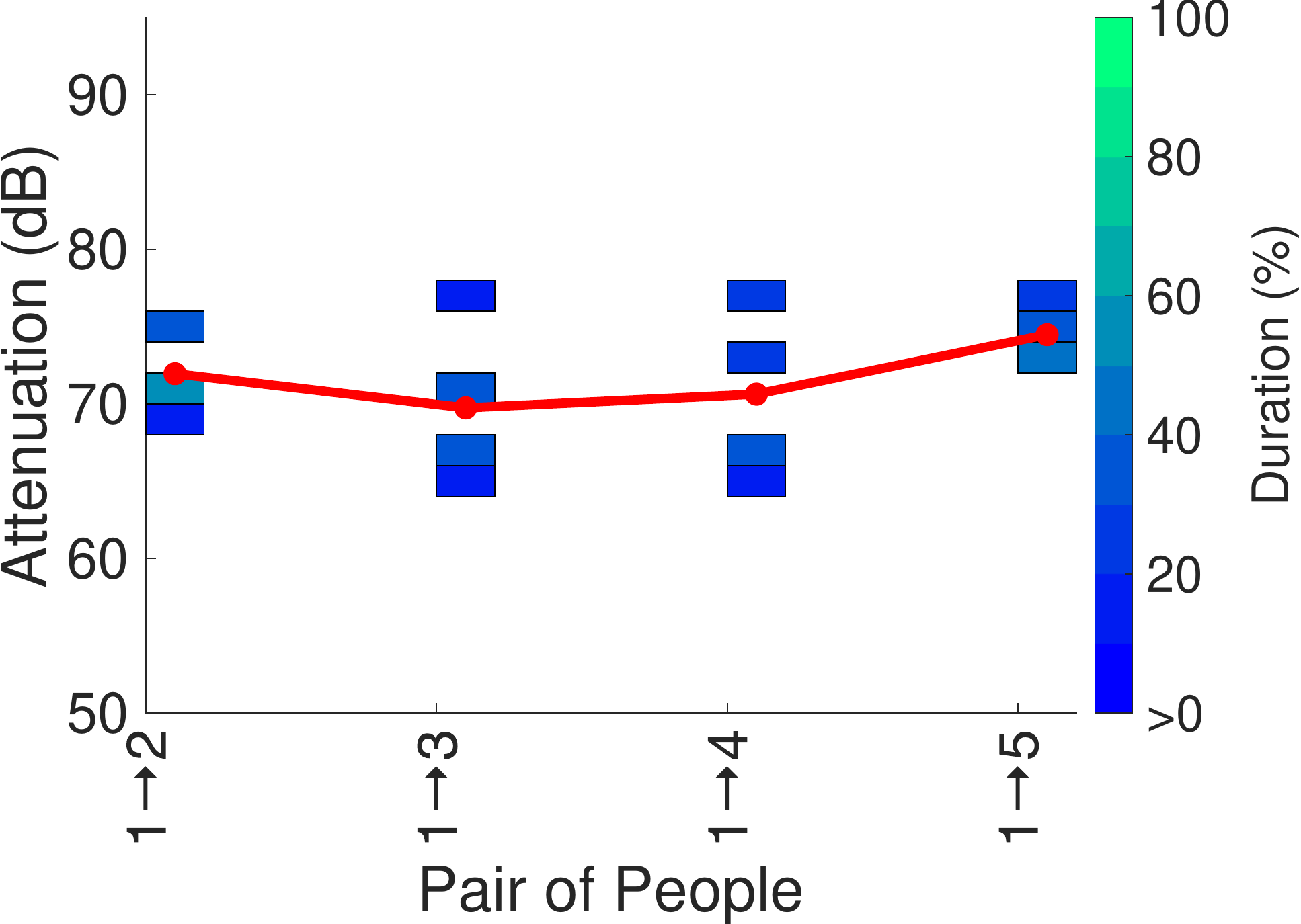}
}
\subfloat[]{
\includegraphics[width=0.45\columnwidth]{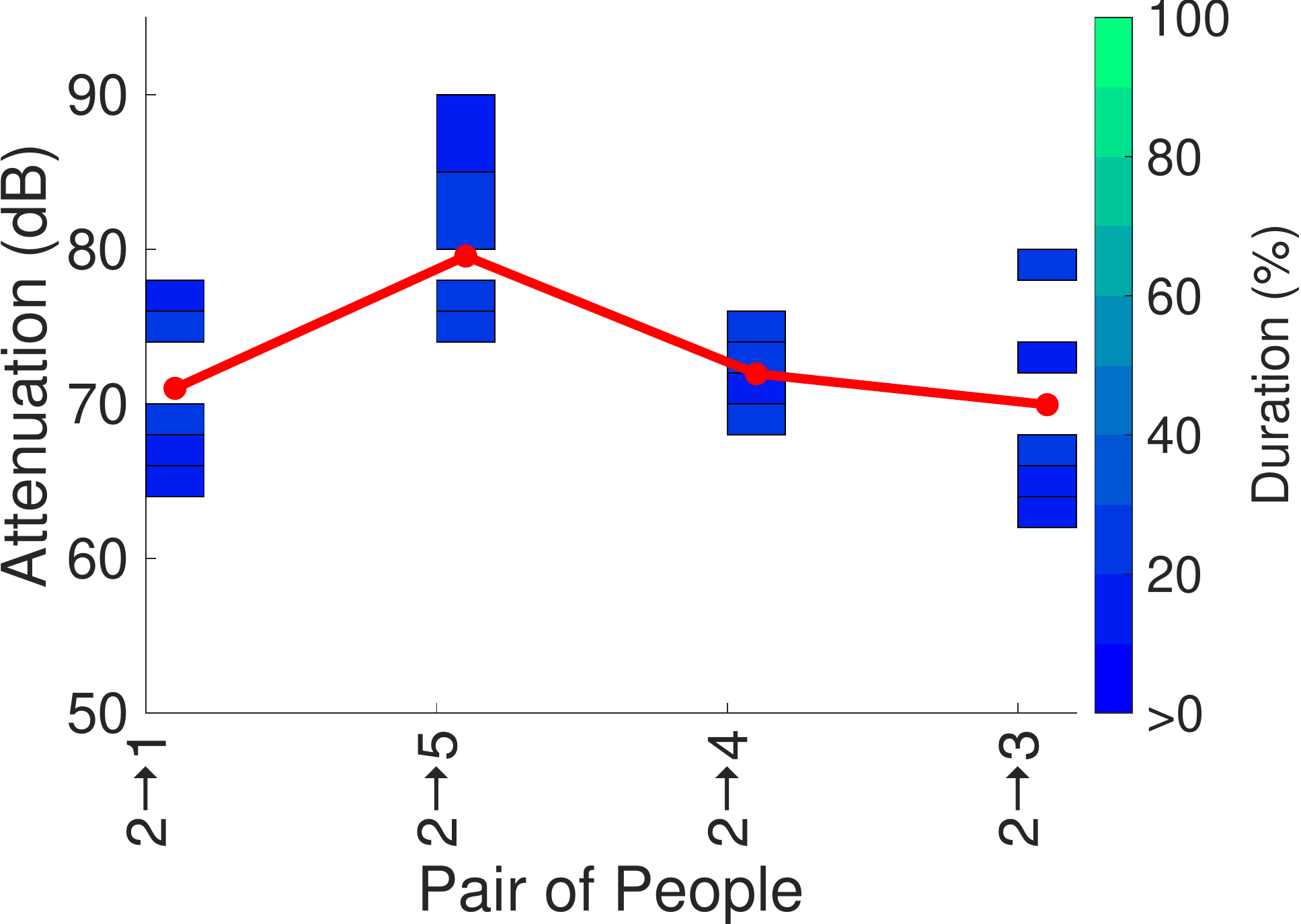}
}\\
\subfloat[]{
\includegraphics[width=0.45\columnwidth]{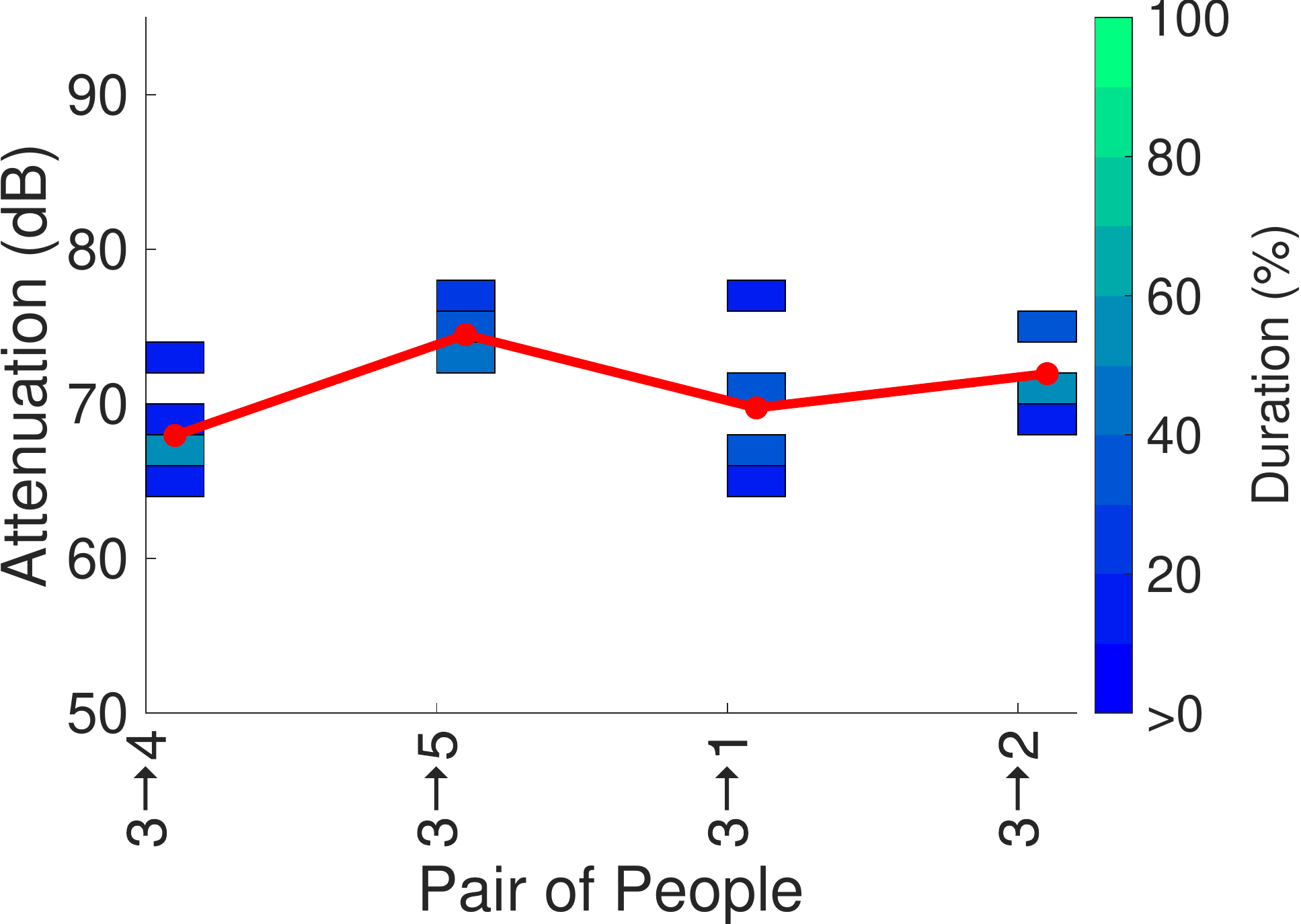}
}
\subfloat[]{
\includegraphics[width=0.45\columnwidth]{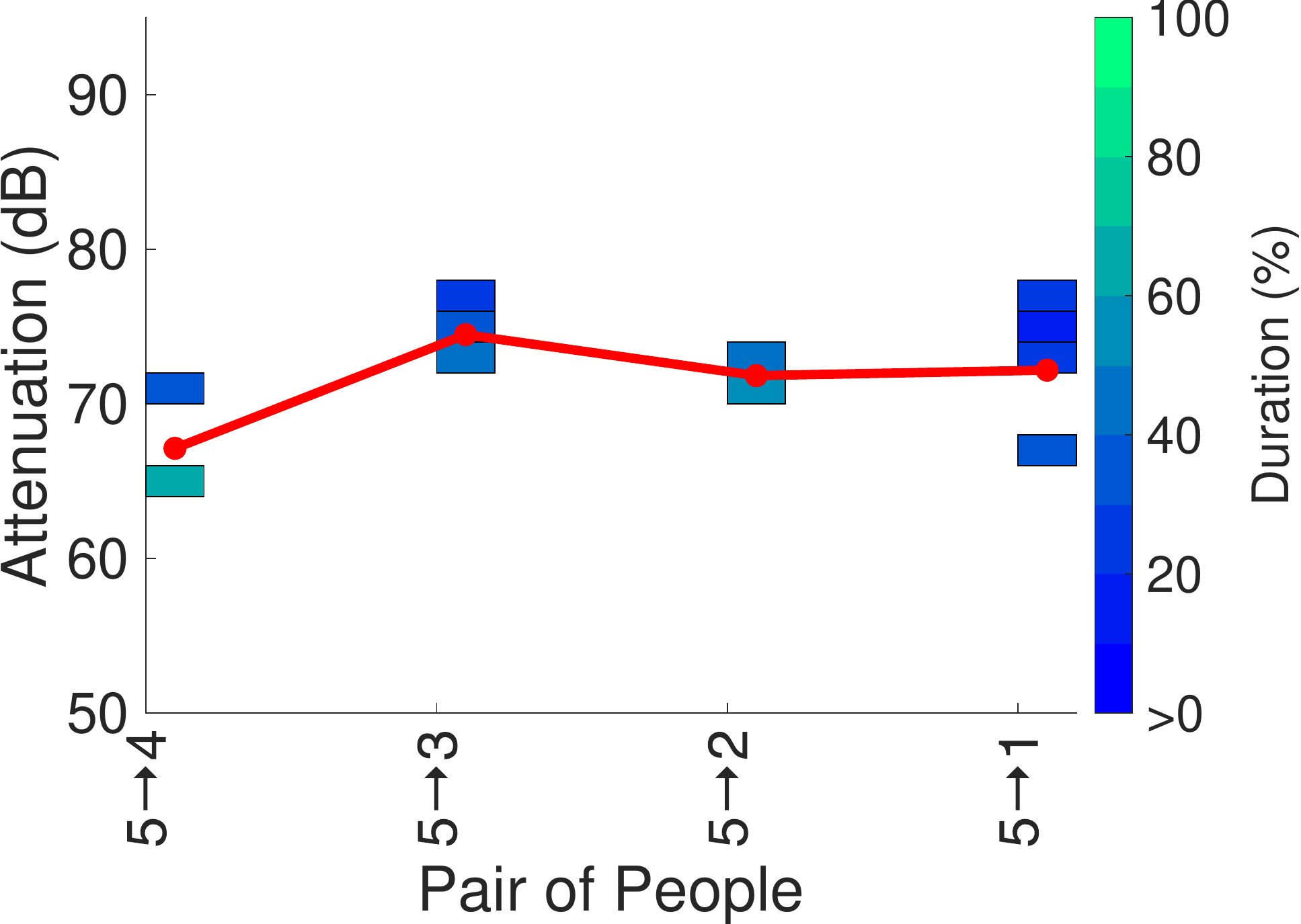}
}
\caption{Attenuation durations reported by GAEN API on completion of the first test on the upper deck of bus.}\label{fig:test3}
\end{figure}

\begin{figure}
\centering
\subfloat[]{
\includegraphics[width=0.45\columnwidth]{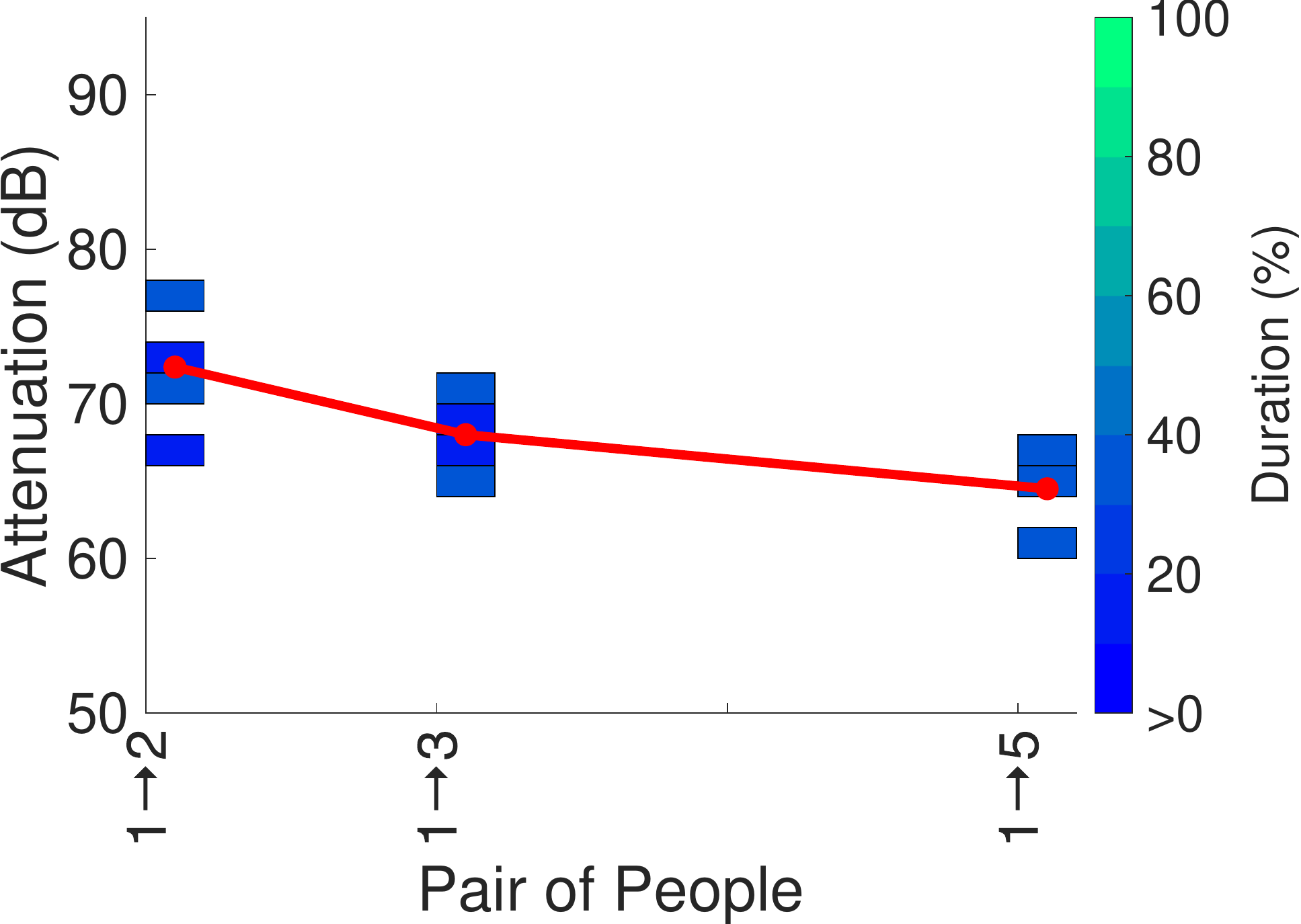}
}
\subfloat[]{
\includegraphics[width=0.45\columnwidth]{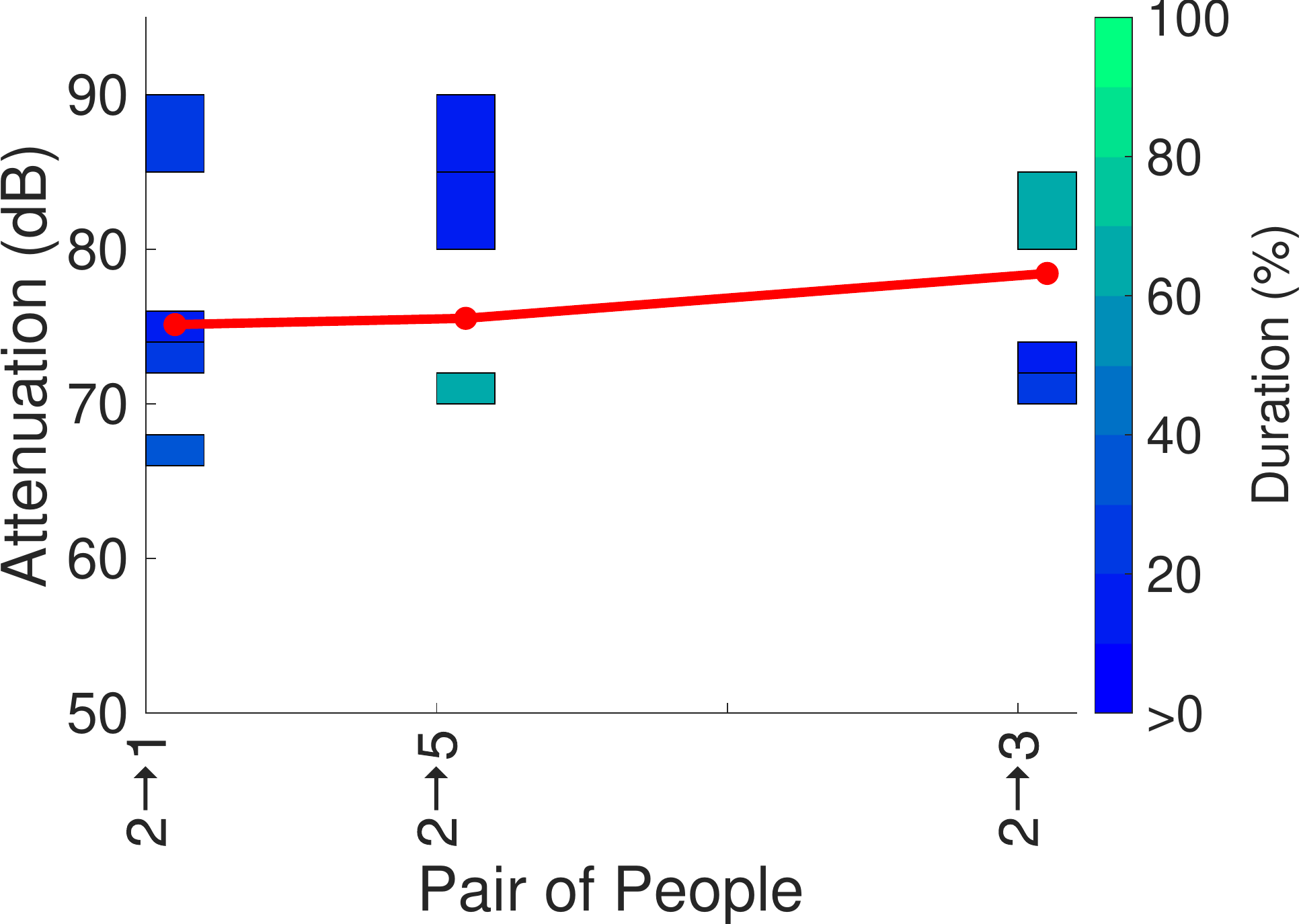}
}\\
\subfloat[]{
\includegraphics[width=0.45\columnwidth]{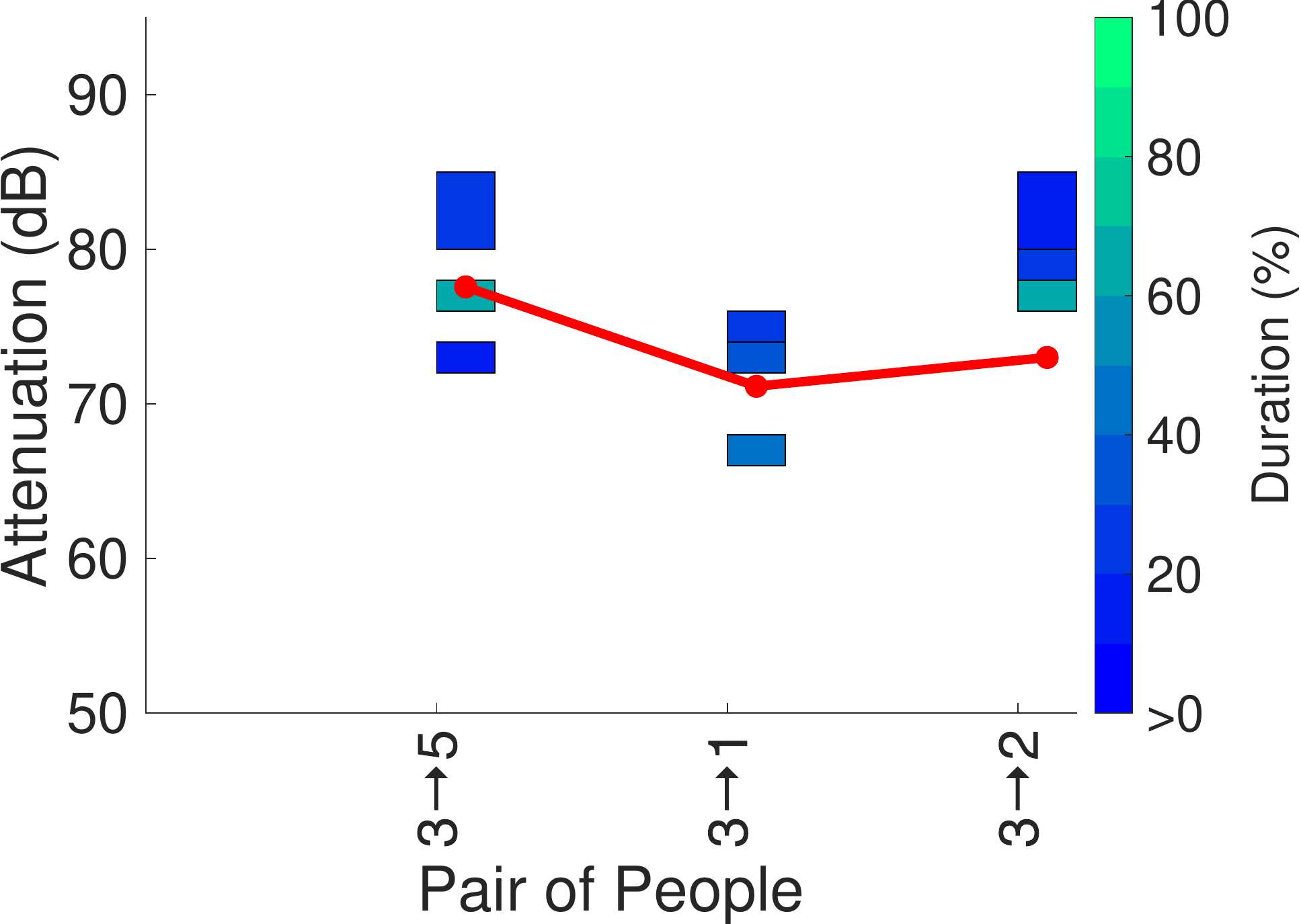}
}
\subfloat[]{
\includegraphics[width=0.45\columnwidth]{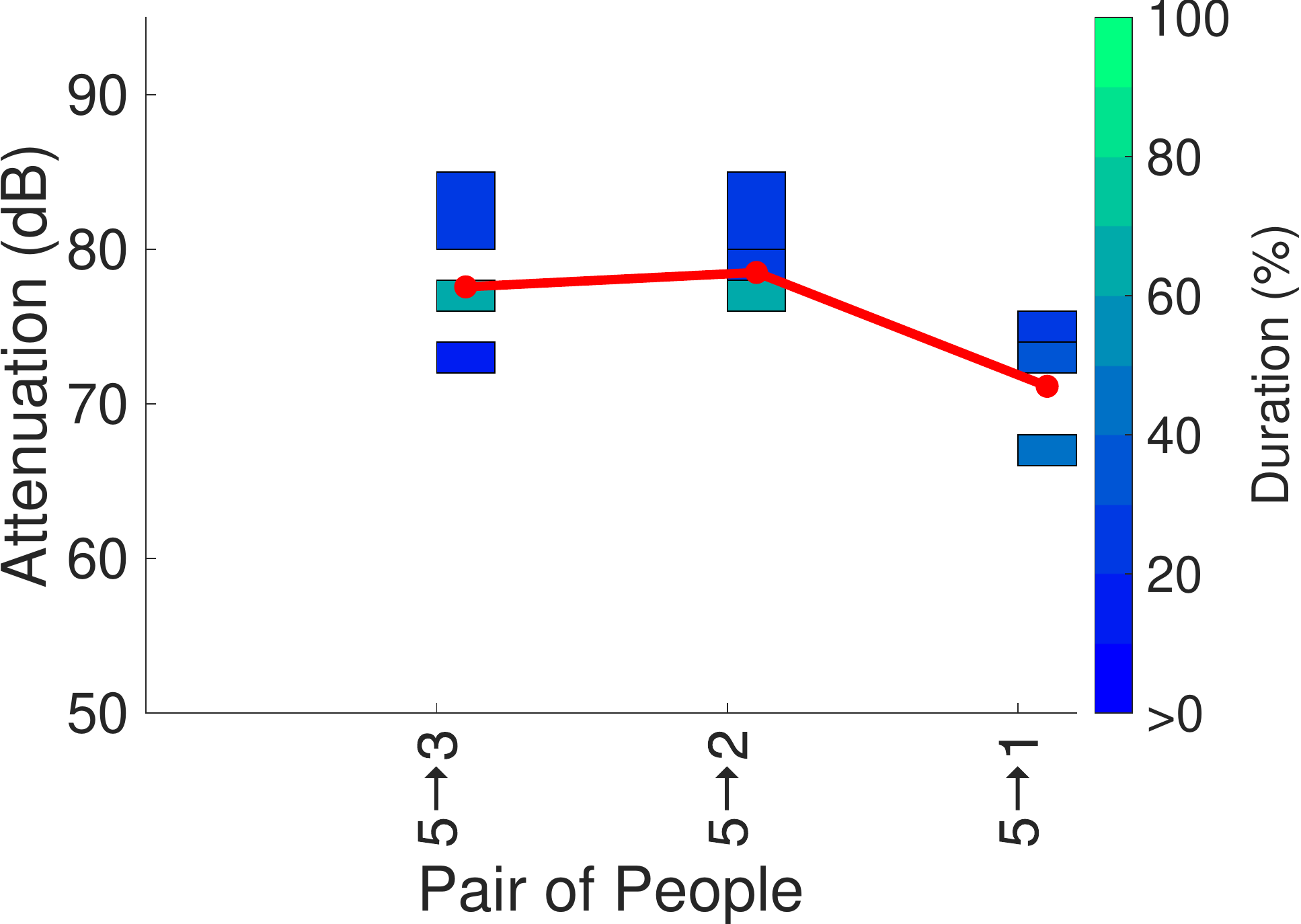}
}
\caption{Attenuation durations reported by GAEN API on completion of the second upper deck test.  Participant in seat 4 is absent for this test but otherwise the participants are the same as in the first test, but with their seating positions swapped about.  We have kept the x-axis labelling the same as in Figure \ref{fig:test3} to facilitate comparison. }\label{fig:test4}
\end{figure}

%%%%%%%%%%%%%%
\subsection{Attenuation vs Distance}\label{sec:dist}
Figure \ref{fig:rssi}(a) plots the attenuation measured between two handsets placed at seat height in the aisle on the upper deck of the bus as the distance between them is varied.   These measurements were taken using the standard Android Bluetooth LE scanner API (rather than the GAEN API).   This scanner API reports an RSSI value for each received beacon.   Following~\cite{duediligencereport2020}, for the Google Pixel 2 handsets used in our experiments we map from RSSI to attenuation level using the formula -17-(RSSI-4).   

It can be seen that the attenuation initially increases as the distance is increased from 0.5m to 1.5m, as might be expected.  But thereafter the attenuation level stays roughly constant with increasing distance (sometimes increasing a little, also sometimes \emph{decreasing} with increasing distance).   The attenuation is around 75dB at 1.5m and also at 3.5m.   

This attenuation behaviour is unusual since generally we expect attenuation to increase with distance.   The floor, ceiling and walls (apart from the windows) of the bus are all made of metal, which is highly reflective at radio frequencies.    We hypothesise that what is happening is that the Bluetooth radio signals are repeatedly reflected from the floor/ceiling/walls and, apart from the signal that escapes out the windows and other smaller apertures, the radio energy is largely conserved as signals travel through the bus.   Whether this is the case or not, however, these baseline measurements indicate that the radio attenuation does not simply with the distance between handsets and this observation is of course pertinent to the use of attenuation level as a proxy for distance.

%%%%%%%%%%%%%%
\subsection{Attenuation Between Passengers}\label{sec:people}
Figures \ref{fig:test1} - \ref{fig:test4} plot the exposure information between handsets reported by the GAEN API for each of the four experiments with seated participants.   

To assist with interpreting the plots the reports in each plot are ordered by increasing distance between the pairs of participants (see Figure \ref{fig:one}(a)).  It can be seen that there is no consistent trend in the change in attenuation level with increasing distance.  Sometimes the attenuation increases with increasing distance (as hoped for when used for proximity detection) but frequently the attenuation level also falls with increasing distance.   This is consistent with the measurements of attenuation vs distance reported in Section \ref{sec:dist}.

Figures \ref{fig:test1} and \ref{fig:test2} both show measurements taken on the lower deck of the bus, but with participants having switched seats between the two.    This allows us to see the impact of differences in the way that each participant uses their handset.   Comparing Figures \ref{fig:test1} and \ref{fig:test2} it can be seen that in plots (b)-(d) the pattern of variation in attenuation is generally similar although the attenuation level can vary substantially with the attenuation level increasing by around 10dB between Figures \ref{fig:test1}(b) and \ref{fig:test2}(b).   Figures \ref{fig:test1}(a) and \ref{fig:test2}(a) differ both qualitatively and quantitatively.   For example, the attenuation between participants 1 and 4 increases by around 10dB from Figure \ref{fig:test1}(a) to \ref{fig:test2}(a) and the attenuation between participants 1 and 2 decreases by around 10dB.  It is difficult to attribute these differences to specific causes, but they do highlight the magnitude of the variation in attenuation that can be induced by person-to-person variation.

Figures \ref{fig:test3} and \ref{fig:test4} show corresponding measurements taken on the upper deck of the bus.   As in the lower deck measurements the general pattern of variation in attenuation is generally similar but there can be changes of around 10dB in the attenuation level, e.g. between participants 1 and 5 in Figures \ref{fig:test3}(a) and \ref{fig:test4}(a), and participants 2 and 3 in Figures \ref{fig:test3}(b) and \ref{fig:test4}(b).

%%%%%%%%%%%%%%%
\subsection{Exposure Notification Error Rate}

The GAEN API is intended for use by health authority Covid-19 contact tracing apps.  When a person is found to be infected with Covid-19 the TEKs from their handset are uploaded to a central server.   The health authority app on another person's handset can then download these TEKs, generate the corresponding RPIs (the values actually sent in beacons) and compare these against the set of RPIs in beacons received by the handset.   If there is a match, the attenuation duration values reported by the GAEN API can then be used to estimate the risk of infection and trigger an exposure notification is this risk is sufficiently high.

A typical requirement is for a person to have spent at least 15 minutes within 2m of the infected person in order to trigger an exposure notification.   The mapping from GAEN attenuation durations to exposure notification is therefore largely based on use of attenuation level as a proxy for proximity between handsets.    

%%%%%%%%%%%%%%%
\subsubsection{Swiss DP-3T Exposure Notification Rule}
Switzerland deployed a Covid-19 contact tracing app based on the GAEN API on 26 May 2020~\cite{swiss_bbc}.  The documentation for this app states that it queries the GAEN API with low and high attenuation thresholds of $t1=50$dB and $t2=55$dB and then bases exposure notifications on the quantity $ES= B1+0.5 B2$, where B1 is the attenuation duration below 50dB reported by the GAEN API and B2 is the attenuation duration between 50dB and $t2$~\cite{swiss}.   An exposure notification is triggered is $ES$ is greater than 15 mins.

With regard to the feasible range of values for $t2$, in ~\cite{techreport2020} measurements are given of RSSI vs distance for Pixel 2 handsets located in an open space outdoors.  Mapping these to GAEN attenuation levels at a distance of 2m the mean attenuation level is 65dB.  Use of $t2$ values significantly above 65dB therefore risks generating a significant number of false positives when used in outdoor environments.

We applied this exposure notification rule to the GAEN attenuation duration dataset reported in Section \ref{sec:people}.   In these experiments all participants are seated within 2m of one another for 15 minutes and so should trigger an exposure notification.   For the 60 pairs of handset locations in this dataset Figure \ref{fig:detect}(a) plots the percentage of these pairs which would trigger an exposure notification as threshold $t2$ is varied from 55dB upwards and the threshold for $ES$ is varied from 5 minutes to 15 mins.   The mean percentage is shown with one standard deviation indicated by the error bars.   The mean and standard deviation are obtained by a standard bootstrapping approach\footnote{The dataset was resampled with replacement $n=1000$ times, the exposure notification percentage calculated for each sample and then the mean and standard deviation of these $n$ estimates calculated.  We selected $n$ by calculating the mean and standard deviation vs $n$ and selecting a value large enough that these were convergent.}   

It can be seen from Figure \ref{fig:detect}(a) that when $t2=55$dB then no exposure notifications are triggered for any choice of $ES$ threshold.   Indeed, when the $ES$ threshold is 10 or 15 minutes no exposure notifications are triggered for any choice of $t2$.   With an $ES$ threshold of 5 minutes the rate of exposure notifications increases with $t2$, as might be expected.  For $t2=65$dB the detection rate is 4\%, rising to 11\% for $t2=68$dB and 31\% for $t2=70$dB.   An $ES$ threshold of only 5 minutes is, however, unrealistic when the medical case requirement is 15 minutes and, as noted above, $t2$ values significantly above 65dB risk generating false positives in outdoor environments.      

%%%%%%%%%%%%%%%
\subsubsection{Threshold-Based Exposure Notification Rule}
\begin{figure}
\centering
\subfloat[Swiss EN Rule]{
\includegraphics[width=0.45\columnwidth]{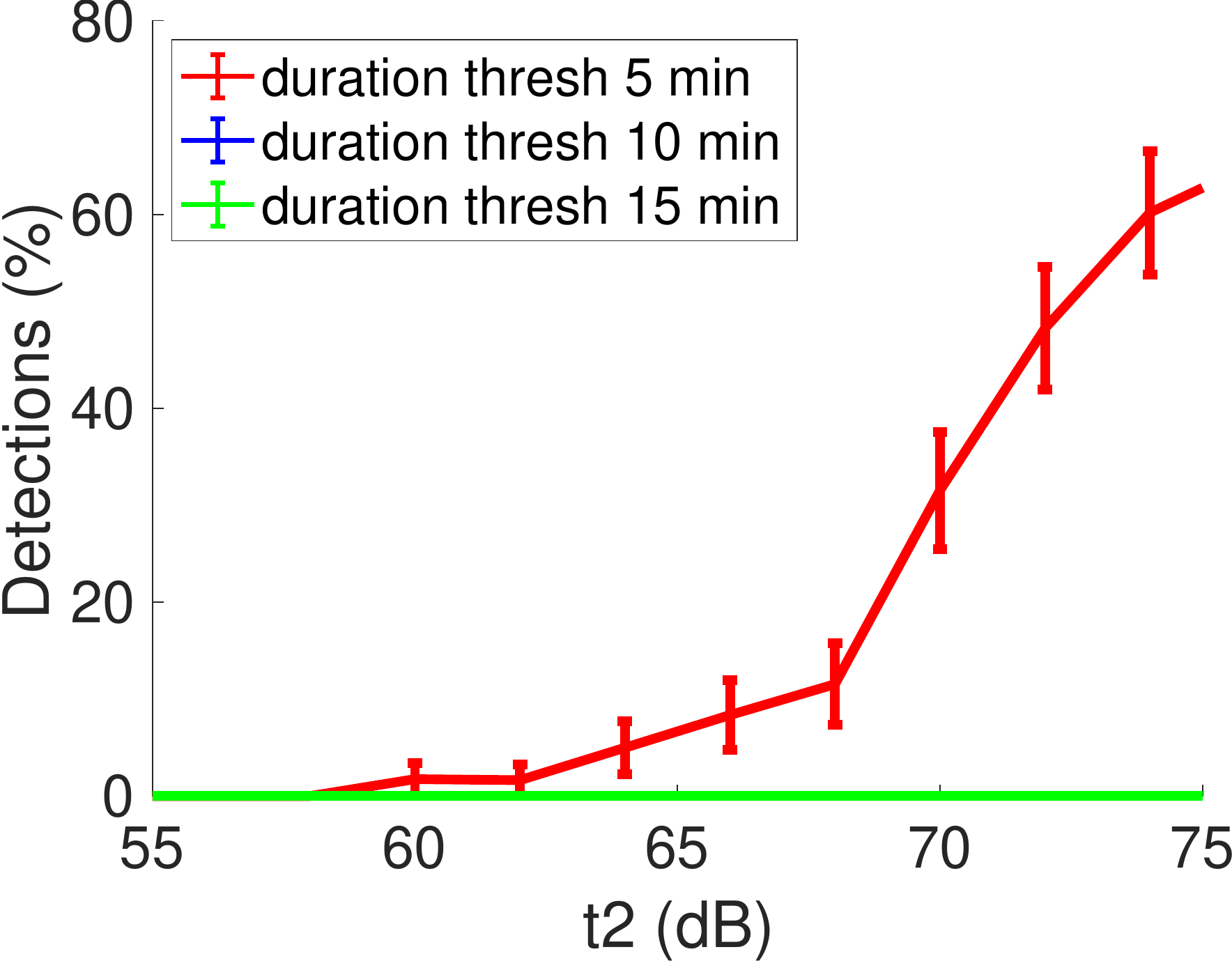}
}
\subfloat[Threshold Rule]{
\includegraphics[width=0.45\columnwidth]{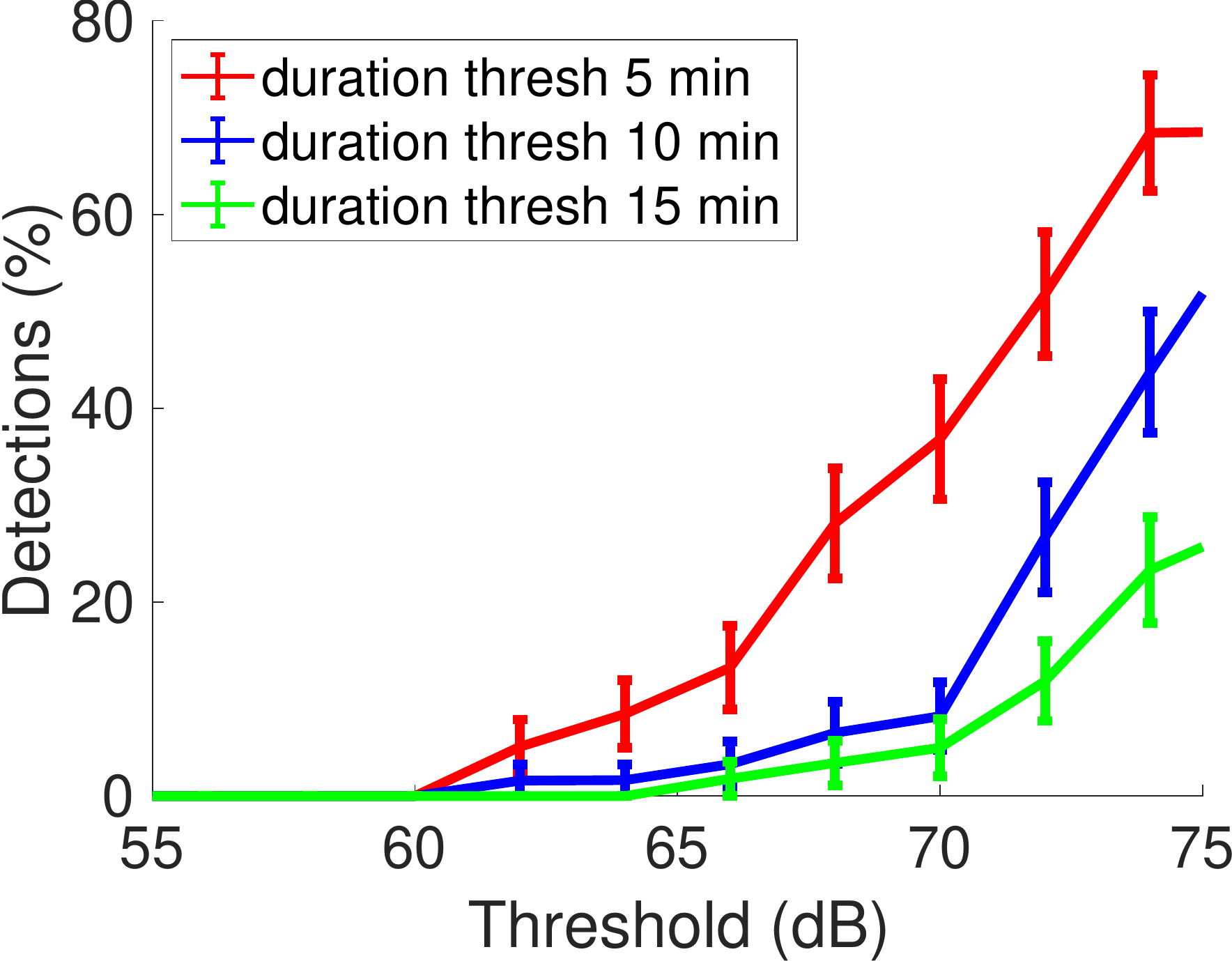}
}
\caption{Exposure notification rate obtained when applying a range of exposure notification rules to the GAEN bus dataset.}\label{fig:detect}
\end{figure}

We also consider the alternative approach of triggering an exposure notification whenever the attenuation duration is above threshold $t2$ i.e. without the weighting of 0.5 used in the Swiss exposure notification rule.   For this exposure notification rule Figure \ref{fig:detect}(b) plots the percentage of exposure notifications as threshold $t2$ is varied from 55dB upwards and the threshold for $ES$ is varied from 5 minutes to 15 minutes. 

It can be seen from Figure \ref{fig:detect}(b) that for $t2=65$dB the detection rate is 1\% when the $ES$ threshold is 15 minutes, 1.5\% for a 10 minute threshold and 10\% for an unrealistic $ES$ threshold of 5 minutes.   Increasing $t2$ to 68dB the detection rates become $3.5\%$, $6.5\%$ and $28\%$ respectively for thresholds of 15, 10 and 5 minutes.  For $t2$ equal to 70dB these figures increase to 5\%, 8\% and 37\%.

\begin{figure}
\centering
\subfloat[]{
\includegraphics[width=0.45\columnwidth]{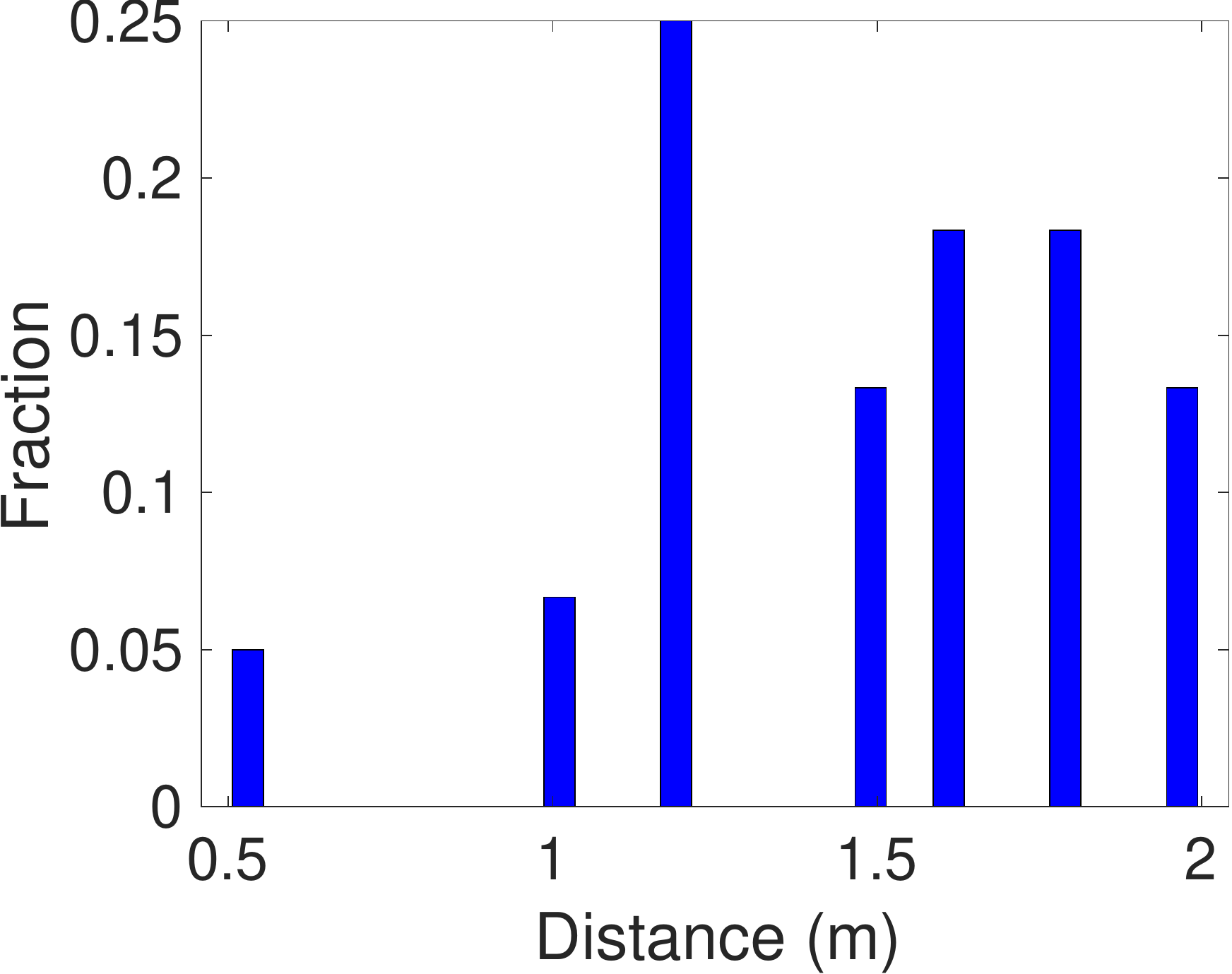}
}
\subfloat[]{
\includegraphics[width=0.45\columnwidth]{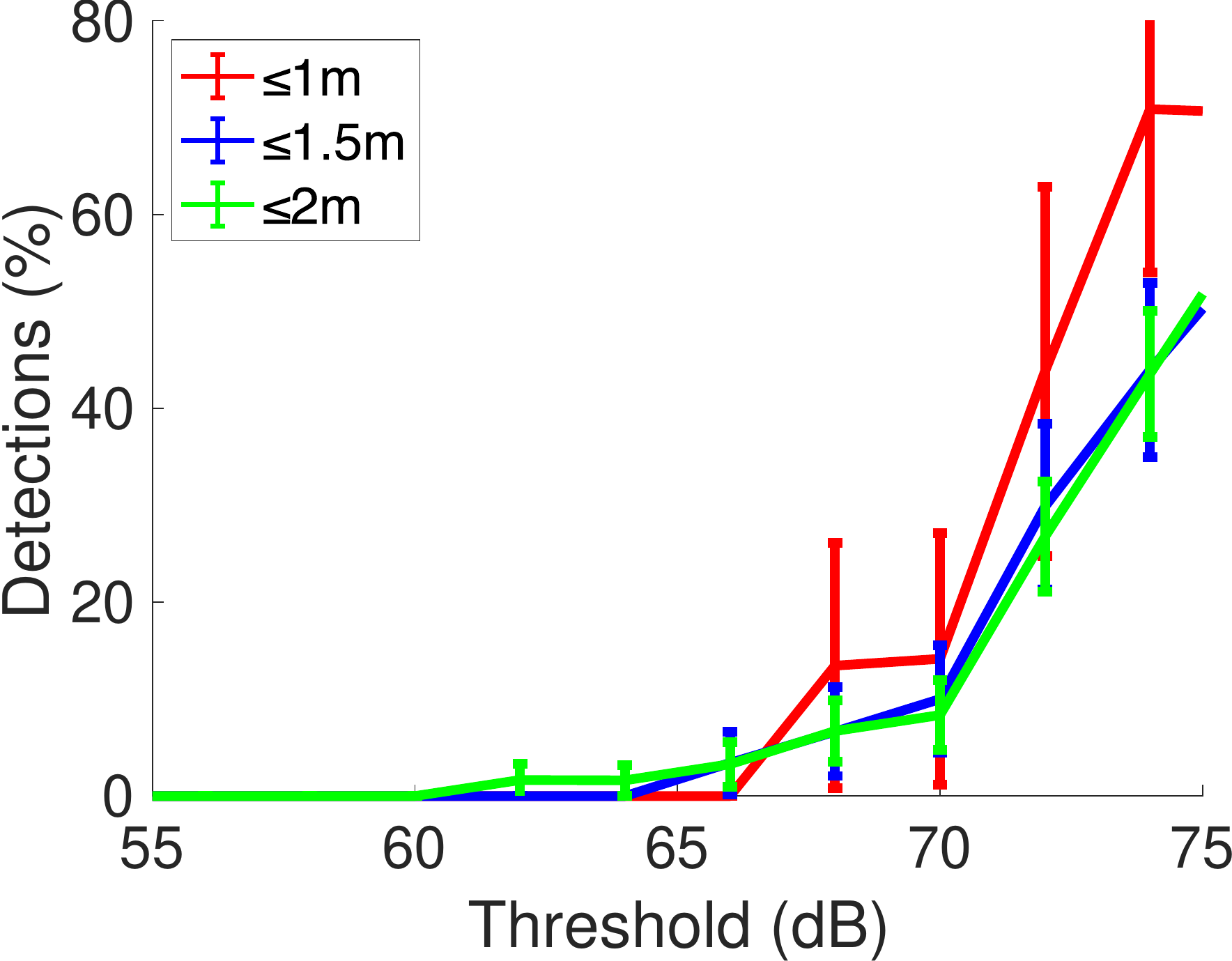}
}
\caption{Exposure notification rate broken down by distance between handsets for the GAEN bus dataset.  Threshold exposure notification rule, $t2$ value marked on x-axis, $ES$ threshold 10 minutes.}\label{fig:detect_dist}
\end{figure}

We also carried out a stratified analysis of exposure notification rates broken down by the distance between handset pairs.    Figure \ref{fig:detect_dist}(a) plots the relative frequencies of distances between the handset pairs in the dataset.  Figure \ref{fig:detect_dist}(b) plots the exposure detection rates vs $t2$ for handsets within 1m, 1.5m and 2m of one another.  The $ES$ threshold is 10 minutes.   It can be seen that the mean detection rate is higher for handsets that are less than 1m apart.   However, for $t2$ values up to 70dB the increase is not statistically significant.    That is, distance between handsets has only a weak, if any, correlation with detection rate.   Further measurements are needed to establish the reason for this, but the baseline data in Figure \ref{fig:rssi}(a) is indicative of the complex radio environment.

%%%%%%%%%%%%%%%
\section{Discussion}
A limitation of this study is that it is confined to handsets using the Android operating system.  The GAEN API is also implemented on Apple iOS devices, but Apple have severely limited the ability of testers to make measurements (each handset is limited to querying the GAEN API a maximum of 15 times a day, and Apple has no whitelisting process to relax this constraint.  Our measurement approach uses 34 queries to extract fine-grained attenuation data per pair of phone locations).

We equipped participants with the same model of handset in order to remove this as a source of variability in the data and instead focus on variability caused by the radio environment and the way that people hold their handsets.   Google and Apple are currently undertaking a measurement campaign to select calibration values within the GAEN API with the aim of compensating for differences between handset models.   We therefore expect that our measurements should also be applicable to a range of handsets, although this remains to be confirmed.

In our experiments we asked each participant to hold the handset in their hand and use it as they usually would when commuting.  Our observations indicate that this is the common case, but it means that we did not collect data for situations where people have the phone in a pocket or bag.  We leave this data collection to future work.   

%\textbf{**handset orientation}

%%%%%%%%%%%%%%%
\section{Conclusion}
We report on the results of a measurement study carried out on a commuter bus in Dublin, Ireland using the Google/Apple Exposure Notification (GAEN) API.   Measurements were collected between 60 pairs of handset locations and are publicly available.  We find that the attenuation level reported by the GAEN API need not increase with distance between handsets, consistent with there being a complex radio environment inside a bus caused by the metal-rich environment.    Changing the people holding a pair of handsets, with the location of the handsets otherwise remaining unchanged, can cause variations of $\pm 10$dB  in the attenuation level reported by the GAEN API.   Applying the rule used by the Swiss Covid-19 contact tracing app to trigger an exposure notification to our bus measurements we find that no exposure notifications would have been triggered despite the fact that all pairs of handsets were within 2m of one another for at least 15 minutes.  Applying an alternative threshold-based exposure notification rule can somewhat improve performance to a detection rate of 5\% when an exposure duration threshold of 15 minutes is used, increasing to 8\% when the exposure duration threshold is reduced to 10 minutes.  Stratifying the data by distance between pairs of handsets indicates that there is only a weak dependence of detection rate on distance.  

%%%%%%%%%%%%%%%
\section*{Acknowledgements}
The authors would like to extend their thanks to the Irish Health \& Safety Executive (HSE) for arranging with Google for us to have whitelisted access to the GAEN API, and to Dublin Bus for kindly providing access to one of their buses.  Trinity College Dublin, (the authors' employer) funded the ``Testing Apps for Contact Tracing'' (TACT) project~\footnote{See \url{https://down.dsg.cs.tcd.ie/tact/}}  that has allowed us the time and handsets required here.   We emphasise that any views expressed in this report are the authors own, and may not be shared by the HSE, Dublin Bus or Trinity College Dublin.  

\bibliographystyle{IEEEtran}
\bibliography{bibfile}
\end{document}